\documentclass[manuscript]{aastex}
\usepackage{color}
\usepackage{ulem}
\slugcomment{Not to appear in Nonlearned J., 45.}

\shorttitle{DISTANCE AND PROPER MOTION MEASUREMENT OF THE RED SUPERGIANT, PZ~CAS, 
IN VERY LONG BASELINE INTERFEROMETRY H$_{2}$O MASER ASTROMETRY}
\shortauthors{Kusuno et al.}

\begin{document}


\title{
  DISTANCE AND PROPER MOTION MEASUREMENT 
  OF THE RED SUPERGIANT, PZ~CAS, 
  IN VERY LONG BASELINE INTERFEROMETRY 
  H$_{2}$O MASER ASTROMETRY
}


\author{K. Kusuno\altaffilmark{1}}
\email{kusuno@vsop.isas.jaxa.jp}

\author{Y. Asaki\altaffilmark{1,2}}
\email{asaki@vsop.isas.jaxa.jp}

\author{H. Imai\altaffilmark{3}}
\email{hiroimai@sci.kagoshima-u.ac.jp}

\author{T. Oyama\altaffilmark{4}}
\email{t.oyama@nao.ac.jp}


\altaffiltext{1}{
    Department of Space and Astronautical Science,
    School of Physical Sciences, 
    The Graduate University for Advanced Studies (SOKENDAI),
    3-1-1 Yoshinodai, Chuou-Ku, Sagamihara, Kanagawa 252-5210, Japan
}
\altaffiltext{2}{
    Institute of Space and Astronautical Science, 
    3-1-1 Yoshinodai, Chuou-Ku, Sagamihara, Kanagawa 252-5210, Japan
}
\altaffiltext{3}{
    Department of Physics and Astronomy, 
    Graduate School of Science and Engineering,
    Kagoshima University, 
    1-21-35 Korimoto, Kagoshima 890-0065, Japan
}
\altaffiltext{4}{
    Mizusawa VLBI Observatory,
    National Astronomical Observatory of Japan,
    2-21-1 Osawa, Mitaka, Tokyo 181-8588, Japan
}


\begin{abstract}
We present the very long baseline interferometry H$_{2}$O maser monitoring observations 
of the red supergiant, PZ~Cas, at 12 epochs from 2006 April to 2008 May. 
We fitted maser motions to a simple model composed of a common annual parallax and 
linear motions of the individual masers.
The maser motions with the parallax subtracted were well modeled by a combination of 
a common stellar proper motion and a radial expansion motion of the circumstellar envelope. 
We obtained an annual parallax of 
0.356~$\pm$~0.026~mas 
and a stellar proper motion of 
$\mu^{*}_{\alpha} \cos{\delta}=-3.7 \pm 0.2$ and 
$\mu^{*}_{\delta}=-2.0 \pm 0.3$~mas~yr$^{-1}$ 
eastward and northward, respectively. 
The annual parallax corresponds to a trigonometric parallax of 
$2.81~^{+0.22}_{-0.19}$~kpc.
By rescaling the luminosity of PZ~Cas in any previous studies using our 
trigonometric parallax, 
we estimated the location of PZ~Cas on a Hertzsprung-Russell diagram and found that it 
approaches a theoretically evolutionary track around 
an initial mass of  $\sim 25$~$M_{\odot}$. 
The sky position and the distance to PZ~Cas are consistent with 
the OB association, Cas~OB5, which is located in a molecular gas super shell. 
The proper motion of PZ~Cas is close to that of the OB stars and other red supergiants 
in Cas~OB5 measured by the {\it Hipparcos} satellite. 
We derived the peculiar motion of PZ~Cas of 
$U_{\mathrm{s}} = 22.8$~$\pm$~1.5, 
$V_{\mathrm{s}} = 7.1$~$\pm$~4.4, and 
$W_{\mathrm{s}} = -5.7$~$\pm$~4.4 km s$^{-1}$. 
This peculiar motion has rather a large $U_{\mathrm{s}}$ component, 
unlike those of near high-mass star-forming regions with negatively large 
$V_{\mathrm{s}}$ motions. The uniform proper motions of the Cas~OB5 member 
stars suggest random motions of giant molecular clouds moving into local potential 
minima in a time-dependent spiral arm, rather than a velocity field caused 
by the spiral arm density wave.  

\end{abstract}





\section{
  Introduction
}\label{sec:01}

Recent very long baseline interferometry (VLBI) astrometry can determine 
distances to Galactic objects on the kiloparsec scale. 
One of thse objects is a red supergiant (RSG) harboring a thick circumstellar envelope.
Since those sources radiate water masers in the radio frequency (RF), 
we can measure trigonometric parallaxes by referring distant extragalactic 
radio sources using a phase-referencing technique (e.g.,
\citealt{Hachisuka2006}). 
If we can measure the trigonometric parallax distance of such a star, 
this enables us to evaluate the absolute magnitude. 
Since the location of an RSG in an evolutionary track of a Hertzsprung--Russell (H-R) 
diagram is not mainly dependent on an effective temperature, 
we can make a constraint on the initial mass from the 
absolute magnitude 
\citep{Choi2008}. 
In addition, because RSGs are sometimes seen in large OB 
associations in the Milky Way, they can become distance indicators for such 
OB associations.  
Determining trigonometric rather than photometric distances to OB 
associations is important for the verification of the calibration method of the 
distance ladder. 

Information about trigonometric parallaxes and proper motions together 
with radial velocities, also enables us to investigate the three-dimensional~(3D) 
motions of the stars in the Milky Way.  
Because proper motions of the water masers around RSGs can be well fitted to 
a simple expanding flow even with some inhomogeneity in it, 
which is different from the case of star-forming regions,
the proper motion of a mass center of an RSG can be rather easily 
estimated from the phase-referencing VLBI. 
\cite{Asaki2010} 
obtained the 3D motion of an RSG, S~Per, and found 
it to have a relatively large deviation (15~km~s$^{-1}$) from the motion expected from 
a flat Galactic rotation curve. 
Considering the age of S~Per is a few tens of million years, it is difficult 
to have a large peculiar motion that old stars often have.
However, it is quite important to know whether or not RSGs' motions are aligned with their 
accompanying OB associations for further discussions about Milky Way dynamics. 
\cite{Asaki2010} 
 pointed out that since S~Per is involved in the Per~OB1 association, such a large 
peculiar motion of an OB association member star may be related to a dynamics 
and mechanism of the spiral arm formation of the Milky Way 
\citep{Baba2009}. 

Here we report on results of phase-referencing VLBI monitoring observations of 
the H$_2$O masers associated with another RSG, PZ~Cas, conducted 
over two years with the VLBI Exploration of Radio Astrometry (VERA) of the National 
Astronomical Observatory of Japan (NAOJ). 
These observations are described in 
Section~\ref{sec:02}. 
Data reduction including astrometric analyses is 
presented in 
Section~\ref{sec:03}. 
Astrometric results of PZ~Cas are presented in 
Section~\ref{sec:04}. 
We discuss the results in 
Section~\ref{sec:05}.  
For 
Sections~\ref{sec:02}--\ref{sec:04}, 
we adopt a solar motion of 20~km~s$^{-1}$ 
\citep{Kerr1986}
relative to the local standard of rest (LSR) 
to the direction of 
($\alpha_{\mathrm{1900}}$,~$\delta_{\mathrm{1900}}$)= 
($18^{\mathrm{h}}$,~$30^{\circ}$). 
Hereafter this solar motion is referred to as the standard solar motion.  
With the standard solar motion, the systemic velocity of PZ~Cas is 
$-36.16$~km~s$^{-1}$, calculated from the heliocentric radial velocity of 
$-45.68\pm0.68$~km~s$^{-1}$ 
\citep{Famaey2005}. 
On the other hand, 
we discuss the observational results on the basis of the solar motion reported by 
\cite{Schonrich2010} 
in 
Section~\ref{sec:05-03}. For the proper motion, we adopt the heliocentric 
coordinates. 

\section{
  Observations
}\label{sec:02}
 
VLBI phase-referencing observations 
of PZ~Cas at 22.2~GHz 
have been conducted at 12 epochs over two years with four VERA antennas at 
Mizusawa, Iriki, Ogasawara, and Ishigakijima 
(coded as MIZNAO20, IRIKI, OGASA20, and ISHIGAKI, respectively). 
The observing epochs are listed in 
Table~\ref{tbl:01}. 

The VERA antennas have a dual beam receiving system each to observe a closely 
located pair of sources simultaneously
\citep{Honma2003}. 
In our monitoring program, one beam (beam~A) observed PZ~Cas and another 
(beam~B) observed a closely located continuum source, 
J233921.1+601011 (hereafter, abbreviated to J2339+6010), 
which was one of the VERA 22~GHz calibrator survey sources 
\citep{Petrov2007} 
$1^{\circ}$.7 away and used as a position reference. 
The observation duration was 8--10 hr. Each observation was divided 
into several sessions for the simultaneous tracking of PZ~Cas and J2339+6010, each 
separated by short-term sessions of bright calibrators (3C~454.3 and 
J230043.0+165514) for 15 minutes in order to check the observing systems.  
We prepared one baseband channel (BBC) with a bandwidth 
of 16~MHz for PZ~Cas and fourteen~BBCs with a total bandwidth of 224~MHz 
for J2339+6010 in left-hand circular polarization. VLBI cross-correlation was carried out with the 
Mitaka FX correlator at NAOJ, Mitaka, Japan, 
to produce VLBI fringe data with 512 spectral channels for PZ~Cas for the central 8~MHz bandwidth, 
corresponding to a velocity spacing of 0.2107~km~s$^{-1}$ for 
the H$_{2}$O $J_{K_{-}K_{+}}=$ $6_{16}-5_{23}$ maser line. 
A typical frequency allocation after the cross-correlation is listed in 
Table~\ref{tbl:02}.

\section{
  Data reduction
}\label{sec:03}

To describe the phase referencing and imaging analysis process, we often refer to the results 
of a specific epoch observation (epoch~F on 2007 March 22) because we started the data analysis 
with this epoch data. 
We analyzed the VLBI data using the standard NRAO data reduction software, 
Astronomical Image Processing Software (AIPS), version 31DEC10. 

\subsection{
  Initial Calibrations
}\label{sec:03-01}

\begin{figure}
\epsscale{0.6}
\plotone{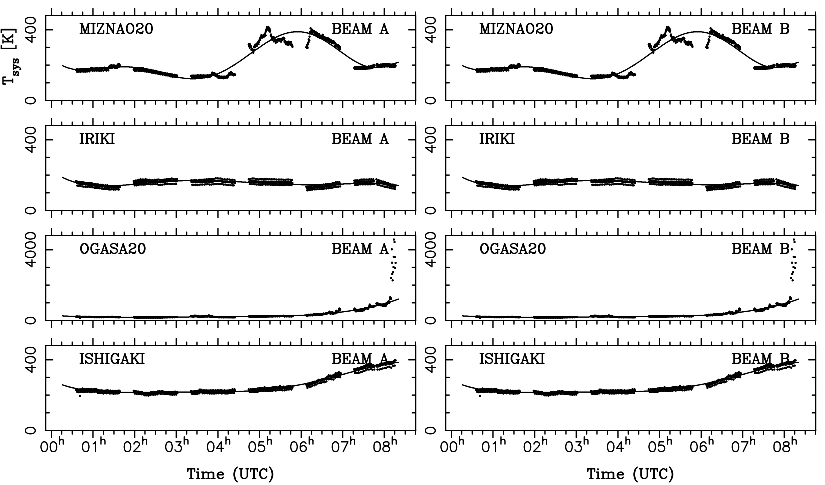}
\caption{
    System noise temperature at epoch F.
    The abscissa is time in UTC and the ordinate is the system noise temperature in kelvin.
    The dotted points are measurements 
    and the solid lines are the polynomial fitting curves after removing the abrupt changes. 
    The left and right panels show beams A and B, respectively. 
    From the top to bottom plots: MIZNAO20, IRIKI, OGASA20, and ISHIGAKI stations. 
}
\label{fig:01}
\end{figure}

Initial calibration tasks were common for both the sources. 
First, the amplitude calibration was carried out by 
using the system noise temperatures ($T_{\mathrm{sys}}$) and gain 
calibration data. The measurements of $T_{\mathrm{sys}}$ often have abrupt 
time variations, as shown in 
Figure~\ref{fig:01}, 
especially when rain falls happened 
during the observation.
We consider such rapid and abrupt time variations to be caused by the 
absorption effect of water puddles on the top of the VERA antenna feedome. 
We carried out a polynomial fitting of the raw $T_{\mathrm{sys}}$ data after 
removing such abrupt changes. 
At epoch~H, since $T_{\mathrm{sys}}$ was not recorded for either beams A or 
B of the ISHIGAKI station. 
Instead, $T_{\mathrm{sys}}$  
of OGASA20 was used for ISHIGAKI, 
because the atmospheric environment (temperature and moisture) is similar.
Note that minor amplitude calibration errors at this stage can be approximately corrected 
in the later self-calibration.
The system noise temperatures for beams A and B were basically approximated to be identical, 
so the time variation of the system noise temperature must be mainly attributed 
to the common atmosphere. 

In the next step, the two-bit sampling bias in the analog-to-digital (A/D) conversion 
in VLBI signal processing was corrected. Earth-orientation-parameter (EOP) 
errors, ionospheric dispersive delays, and tropospheric delays were calibrated 
by applying precise delay tracking data calculated with an updated VLBI correlator model 
supplied by the NAOJ VLBI correlation center. 
The instrumental path differences between the two beams were corrected using 
the post-processing calibration data supplied by the correlation center 
\citep{Honma2008}. 
The phase tracking centers for PZ~Cas and J2339+6010 of the updated correlator 
model were set to the positions given in 
Table~\ref{tbl:03} 
for all the observing epochs. 

\subsection{
  Phase Referencing and Imaging
}\label{sec:03-02}

\begin{figure}
\epsscale{0.33}
\plotone{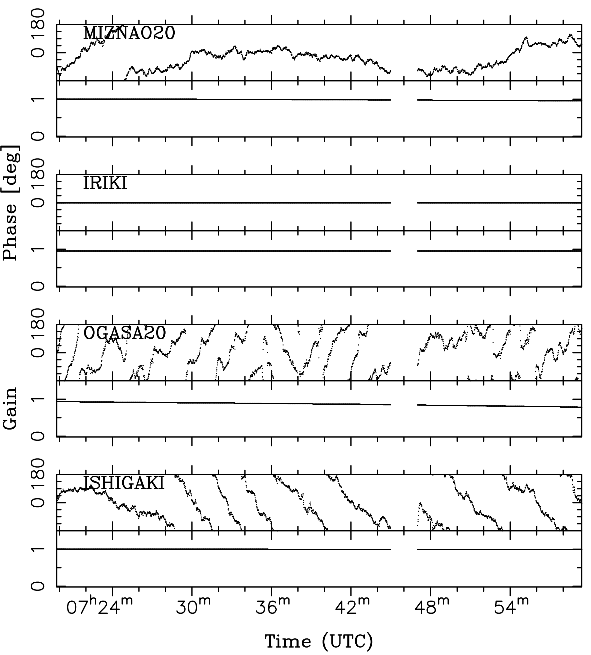}
\plotone{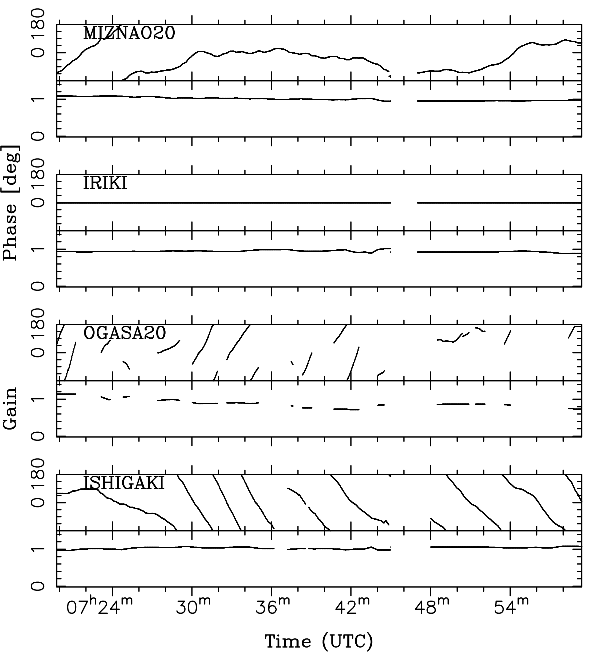}
\caption{
    Complex gain calibration solutions obtained from J2339+6010 at epoch~F for the phase referencing. 
    IRIKI is the reference antenna. Each antenna 
    has two plots: the abscissa is time in UTC, and the ordinate is the phase calibration data 
    in degree 
    and the amplitude gain 
    for the upper and lower plots, respectively. 
    The left and right plots show the direct phase-transfer (DPT) solutions introduced in Section~\ref{sec:03-02}
    and those obtained in an AIPS phase-referencing analysis (AIPS CL table), respectively. 
    For making the AIPS CL table, the solution intervals were set to 2, 2 and 10 minutes 
    for the fringe fitting, phase-only self-calibration, and amplitude-and-phase self-calibration, 
    respectively. 
    From the top to bottom plots: MIZNAO20, IRIKI, OGASA20, and ISHIGAKI. 
}
\label{fig:02}
\end{figure}

After the initial calibration, an image of the reference source, J2339+6010, was synthesized using 
the standard VLBI imaging method with the fringe fitting and self-calibration. 
We carried out the above process twice to 
make the image of J2339+6010: the first process started with a single point source model 
and the second with 
the reference source image that had been made in the previous process. 
After imaging, the phase-referencing calibration solutions were directly calculated  
from the reference source fringe data: the calibrating phase, 
$\Phi_{\mathrm{cal}}(t)$, at time $t$ was 
obtained from the following calculation with a vector average of all the spectral channels 
in all the BBCs: 
\\
\begin{eqnarray}
\mathrm{exp}
[
i \Phi_{\mathrm{cal}}(t)
]
 &=& 
\sum^{N}_{n=1}
  \sum^{M}_{m=1}
    \mathrm{exp}
    \{
      i [ \Phi^{\mathrm{r}}_{\mathrm{raw}}(n, m, t)  \nonumber \\
  &+& 2\pi [\nu^{\mathrm{r}}_{n} + (m-1)\Delta \nu - \nu^{\mathrm{t}}_{0}]
               \Delta\tau^{\mathrm{r}}_{\mathrm{g}}(t)  \nonumber \\
  &-& \Phi^{\mathrm{r}}_{\mathrm{v}}(U(t), V(t))]
    \}, \nonumber
\end{eqnarray}
\\
where
$\Phi^{\mathrm{r}}_{\mathrm{raw}}$ is the raw data fringe phase of the reference source, 
and 
$\Phi^{\mathrm{r}}_{\mathrm{v}}(U(t), V(t))$ is the visibility phase calculated from 
the CLEAN components of the reference source image for a baseline 
with $U(t)$ and $V(t)$.  
$N=14$ is the BBC number, 
$M=64$ is the spectral channel number in the single BBC, 
$\nu^{\mathrm{t}}_{\mathrm{0}}$ is the RF frequency of the strongest H$_{2}$O maser emissions 
in beam~A, 
$\nu^{\mathrm{r}}_{n}$ is the RF frequency at the lower band edge of the $n$th BBC 
for the reference source as listed in 
Table~\ref{tbl:02}, 
$\Delta \nu=250$~kHz is the frequency spacing in the single BBC for the reference source, 
$m$ is the integer number indicating the order of the frequency channel in the single BBC, 
and $\Delta \tau^{\mathrm{r}}_{\mathrm{g}}(t)$ is the third-order polynomial fitting result of the temporal 
variation of the group delay obtained 
from the preceded fringe fitting for the reference source. 
We refer to the phase calibration solutions obtained with this method as direct phase-transfer~(DPT) 
solutions. The DPT solutions were then smoothed by making a running mean with an averaging time 
of 16~s in order to reduce the thermal noise. 
The smoothed DPT solutions, as well as the one obtained with an ordinary AIPS phase-referencing analysis 
(AIPS CL table), are shown in 
Figure~\ref{fig:02}. 
Those two results are basically consistent with each other. 
Note that even when solutions could not be obtained with 
the fringe fitting or the following AIPS self-calibration because the signal-to-noise ratios (S/N) 
was below a cutoff value, the DPT solutions were created from the raw fringe data, 
as shown for the OGASA20 station in 
Figure~\ref{fig:02}. 
The final astrometric result with the smoothed DPT solutions is consistent 
with the AIPS CL table at epoch~F.

Amplitude gain adjustment was performed using a third-order polynomial fitting for 
the amplitude self-calibration solutions for J2339+6010. Since S/N of J2339+6010 was 
unexpectedly low at epoch G, we could not obtain good solutions for J2339+6010 in the fringe-fitting. 
We therefore obtained the group delay solutions from 3C~454.3 and set the amplitude gain 
to the unity for all the observing time. The DPT solutions were 
created from the J2339+6010 fringe data together with the group delay solutions 
determined with 3C~454.3. At epoch~I, since IRIKI did not attend the 
observation, we used the remaining three stations in the data reduction. 

\begin{figure}
\epsscale{.35}
\plotone{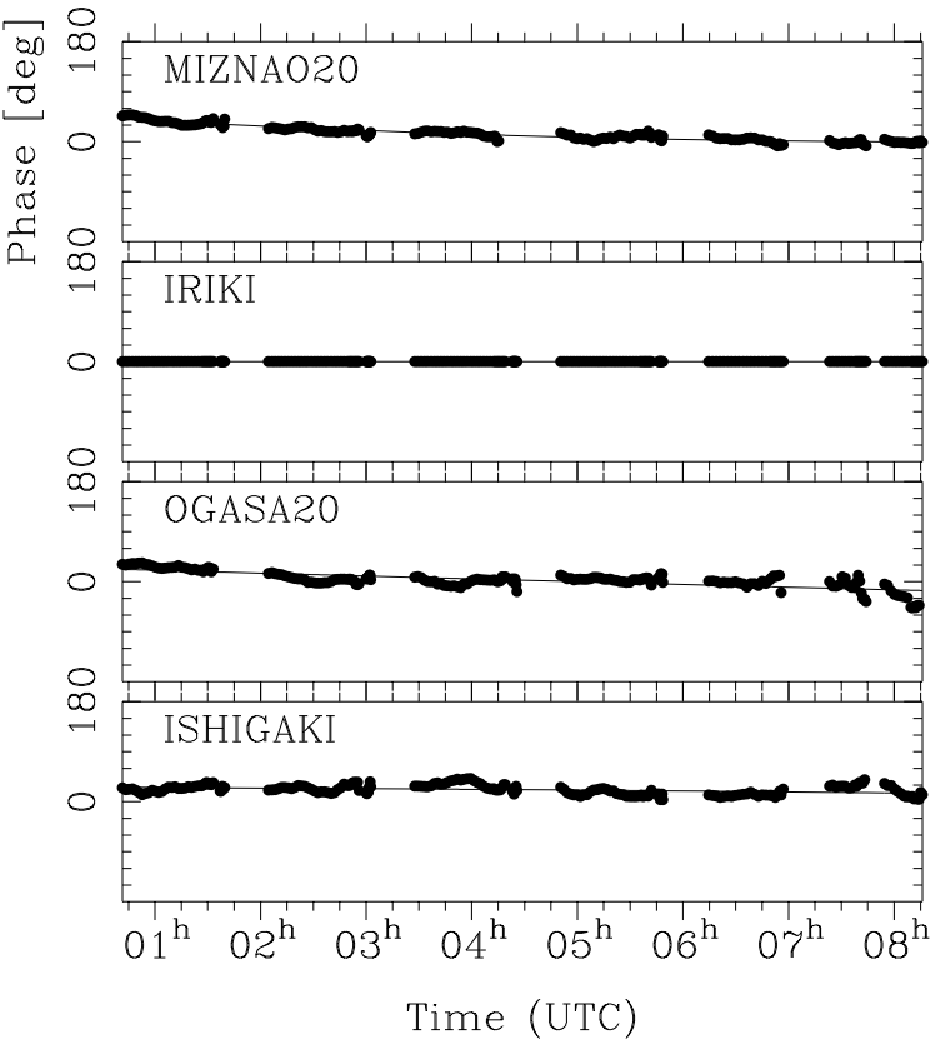}
\plotone{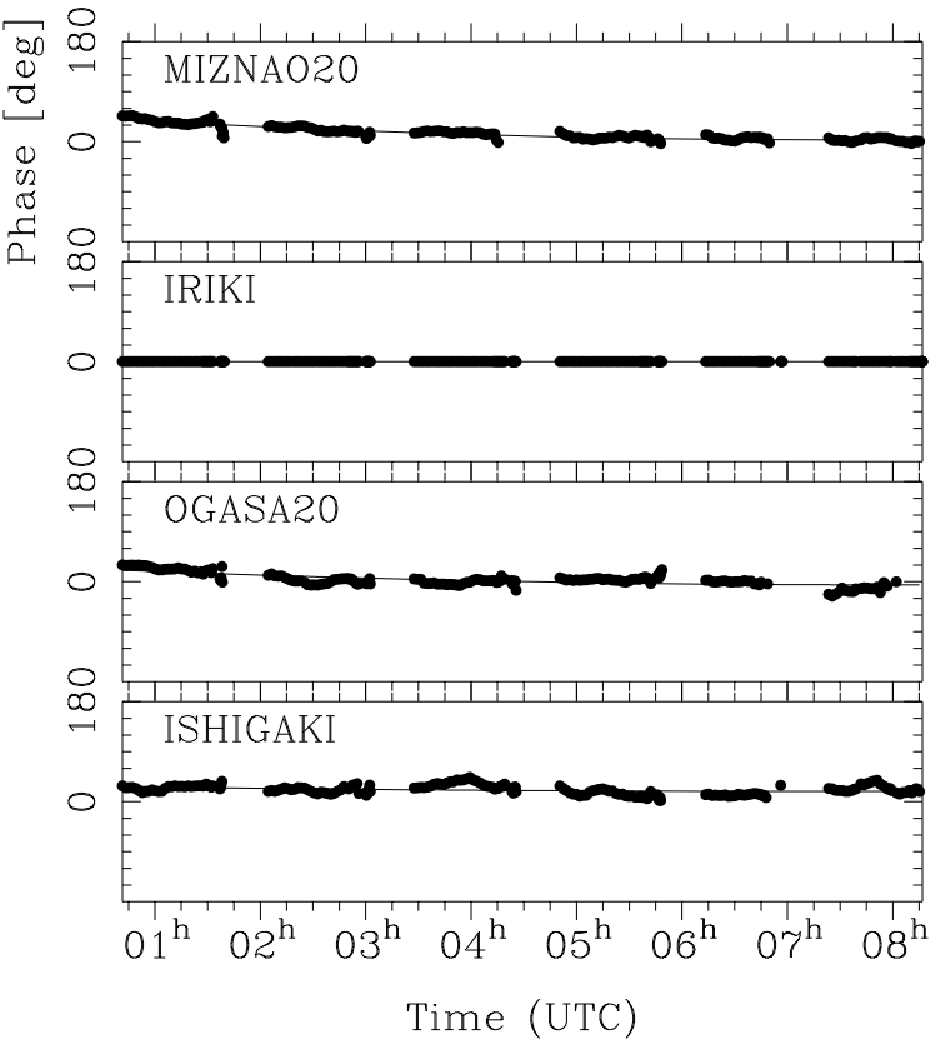}
\caption{
    Phase solutions using the fringe fitting with a solution interval of 20~minutes 
    for the PZ~Cas's brightest velocity channel 
    ($V_{\mathrm{LSR}} =-42.0$~km~s$^{-1}$) at epoch~F.  
    The abscissa is the time in UTC, and the ordinate is the phase in degree. 
    IRIKI is the reference antenna. 
    The dotted points are the fringe-fitting solutions, and the solid lines represent 
    fitting results with a third-order polynomial for each station. 
    The left and right plots show the fringe fitting results after the phase referencing 
    with the direct phase-transfer calibration and the AIPS CL table, respectively. 
    From the top to bottom plots: MIZNAO20, IRIKI, OGASA20, and ISHIGAKI. 
}
\label{fig:03}
\end{figure}

The Doppler shift in PZ~Cas spectra 
due to Earth's rotation and revolution was corrected after the phase referencing. 
The observed frequencies of the maser lines were converted to radial velocities 
with respect to the traditional LSR using the rest frequency of 22.235080~GHz for the 
H$_2$O $6_{16}-5_{23}$ transition. 
Since the cross power spectra within the received frequency region showed a 
good flatness both in the phase and the amplitude, the receiver complex gain 
characteristics were not calibrated in this analysis. 
We inspected the cross power spectra for all the baselines in order to select 
a velocity channel of the maser emission which would be unresolved and not 
have rapid amplitude variation. 
At this step, fast phase changes with a typical timescale of a few minutes to 
a few tens of minutes can be removed by the phase referencing, 
but slow time variations including a bias component as shown in 
Figure~\ref{fig:03} 
cannot. 
The maser image was distorted because of the residual time 
variations of the fringe phase on a timescale of several hours, 
which might be caused by uncertainties in the updated VLBI correlator model. 
We conducted the fringe fitting for the selected velocity channel 
with a solution interval of 20~minutes. We fitted the solutions to a third-order 
polynomial, as shown in 
Figure~\ref{fig:03}, 
and removed it from both of the fringe phases of PZ~Cas and J2339+6010. 
This procedure removes the residual fringe phase error from the PZ~Cas data 
while it is transferred to the J2339+6010 data. 
As a result, this leads to 
an improvement of the PZ~Cas image quality, but it also creates a distortion of the 
J2339+6010 image. This distortion of the reference source causes an astrometric 
error in the relative position between the sources, as later discussed in Section~\ref{sec:03-03}. 

Figure~\ref{fig:03} 
also shows the fringe phase residuals obtained with the AIPS CL table. 
Although the fringe fitting results with the AIPS CL table can be almost identifiable to 
those with the DPT solutions, it is later noted that the image peak flux density at 
$V_{\mathrm{LSR}}=-42.0$~km~s$^{-1}$ with the DPT solutions 
is 2\% lower than that with the AIPS CL table at epoch~F. 
In addition, the image root-mean-square noise 
resulting from the DPT solutions is 5\% higher than that resulting 
from the AIPS CL table.  
We can also see that the residual phases resulting from the 
DPT solutions in the fifth and sixth observing sessions in 
Figure~\ref{fig:03} 
have a larger deviation than
those resulting from the AIPS CL table. Although 
the number of the calibrated 
visibilities after applying the DPT solutions was larger than that 
with the AIPS CL table, especially in the last two sessions, the 
visibilities with lower S/N might degrade the final image quality. 
We have to admit that the AIPS CL table may lead to a slightly higher quality 
in the image synthesis 
because the fringe fitting and self-calibration take off visibility with low S/N. 
The problem is that there were some observations whose data 
showed unexpectedly low S/N for certain baselines. 
For such cases, it was hard to create the AIPS CL table 
with the fringe fitting and self-calibration. For the following three reasons, we 
used the DPT solutions instead of the AIPS CL table for all the observing epochs: 
(1) the DPT is useful even when the fringe fitting fails because 
of the low S/N, 
(2) the final astrometric results between the two methods have little difference, and 
(3) it is preferable to use a unified analysis scheme for obtaining scientific results from 
a data set with the same ($U$,~$V$) distribution. 

The DPT solutions and the long-term phase calibration data were 
applied to the remaining frequency channels to make image cubes   
for the LSR velocity range of $-23.0$ to $-60.8$~km~s$^{-1}$ with 
a step of 0.2107~km~s$^{-1}$. 
Image synthesis with 60~$\mu$as pixel for a region of $4096\times4096$ pixels, 
roughly a $246\times246$ square milliarcsecond (mas) region, was made by using IMAGR. 
To initially pick up maser emissions in the maps, 
we used the AIPS two-dimensional (2D) Gaussian component survey task, SAD, 
in each velocity channel map in order to survey emissions. 
We visually inspected the surveyed Gaussian components one by one. 
We stored the image pixels that were larger than a 5$\sigma$ noise level for a 
visually confirmed maser emission. 
Hereafter we define a maser ``spot'' as an emission in a single velocity channel, and 
a maser ``feature'' as a group of spots 
observed in at least two consecutive velocity channels at a coincident 
or very closely located positions. 

\subsection{
  Accuracy of the Relative Position of PZ~Cas
}\label{sec:03-03}

In the phase-referencing astrometric analysis, positions of the PZ~Cas maser spots 
were obtained with respect to that of the J2339+6010 image. 
The positional error includes the effect of a peak position shift 
due to uncertainties in the VLBI correlator model 
(atmospheric excess path lengths [EPLs], antenna positions, EOP, and so on) for the sources. 
Hereafter we refer to the relative position error of 
a pair of point sources using the phase referencing due to uncertainties 
in a VLBI correlator model as an astrometric error. 

\begin{figure}
\epsscale{.35}
\plotone{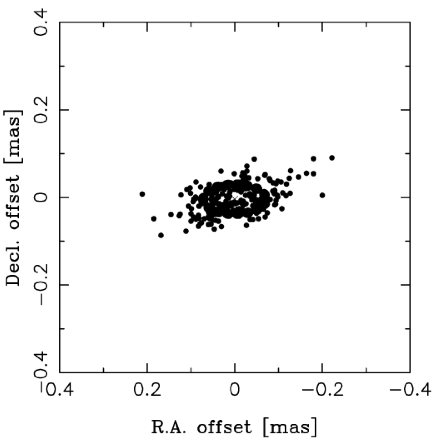}
\plotone{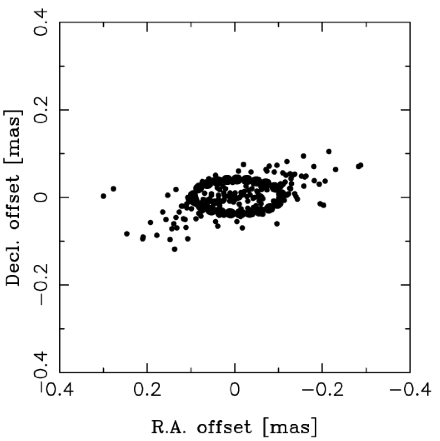}
\caption{
    Astrometric observation simulation results for the pair of PZ~Cas and J2339+6010. 
    The dots represent PZ~Cas's image peak positions relative to J2339+6010 at 22~GHz 
    with the VERA for 200 trials. The 
    ellipses represent 1$\sigma$ for the distributions. 
    Left: the VERA full array case. 
    Right: the VERA three-station case (MIZNAO20, OGASA20, and ISHIGAKI). 
}
\label{fig:04}
\end{figure}

We conducted Monte Carlo simulations of phase-referencing observations for 
the pair of PZ~Cas and J2339+6010 to estimate the astrometric error. 
For generating simulated phase-referencing fringes, 
we used an improved version of a VLBI observation simulator, ARIS 
\citep[Astronomical Radio Interferometer Simulator;][]{Asaki2007}, 
with new functions simulating station clocks 
\citep{Rioja2012} 
and amplitude calibration errors.   
In the simulations, we assumed flux densities of PZ~Cas and J2339+6010 
of equivalently 10~Jy and 0.2~Jy for 15.6~kHz and 224~MHz bandwidths, 
respectively, which were determined from our observations. 
The sources were assumed to be point sources. The observation schedule and 
observing system settings in the simulations, such as the recording 
bandwidth and A/D quantization level, were adopted from the actual VERA observations. 
A tropospheric zenith EPL error of 2~cm and other parameters were set as 
suggested for VERA observations 
\citep{Honma2010}.
For simplification of the simulations, reference source 
position errors were not considered. The simulation results from 200 trials are 
shown in 
Figure~\ref{fig:04}. 
They show that the 1$\sigma$ errors in the relative 
position are 72 and 33~$\mu$as  
in right ascension and declination, respectively. 
We used these standard deviations as the astrometric 
error in the following analysis. 

At epoch~I, the IRIKI station geographically located 
at the center of the VERA could not attend the observation, so the astrometric 
accuracy may become worse with the remaining three stations. 
We conducted another simulation series for the three-station case, 
and the resultant astrometric accuracy is 101 and 39~$\mu$as for right ascension 
and declination, respectively. We therefore used this astrometric accuracy  
at epoch~I in the following analysis.

\section{
  Results
}\label{sec:04}

\begin{figure}
\epsscale{.60}
\plotone{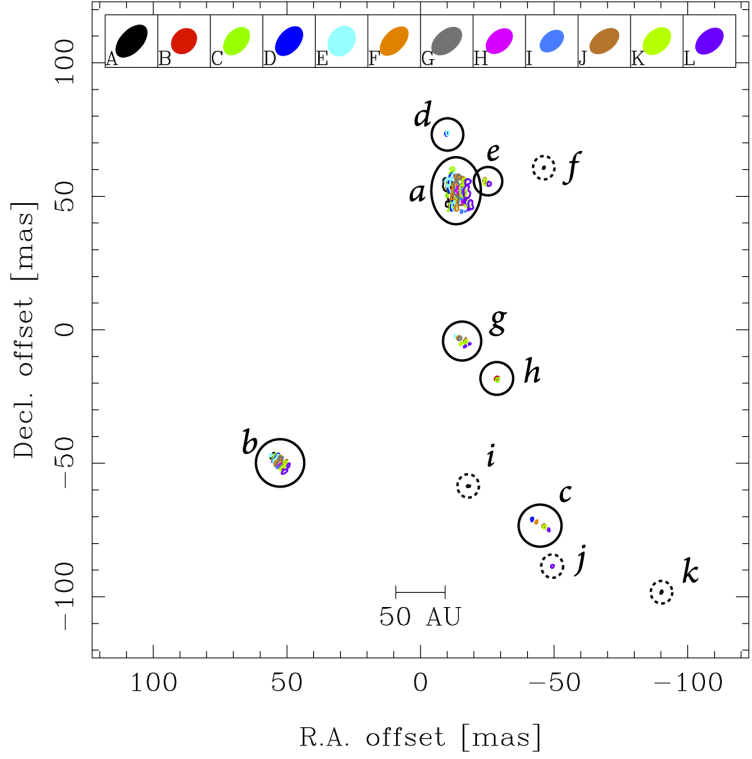}
\caption{
    Spatial distribution of the PZ~Cas H$_{2}$O masers in the 12 epochs. 
    The outlines of maser features represent a 5$\sigma$ noise-level 
    contour (418, 722, 661, 367, 467, 344, 309, 658, 686, 440, 313, and 285~mJy 
    at epochs A, B, C, D, E, F, G, H, I, J, K and L, respectively). 
    The synthesized beams are shown in the upper 1$\times$1 square mas boxes. 
    Maser features surrounded with circles represent maser groups.  
    Groups $\it {~f,~i,~j}$, and~${k}$ are features detected at a single epoch.
    Group $\it {a}$ is composed of four features. Groups $\it {b}$ and $\it {g}$ are 
    composed of two features. The map origin is the phase-tracking center.
      }

\label{fig:05}
\end{figure}

Maser features of PZ~Cas are spatially distributed over a region of 200 $\times$ 200 square mas. 
We identified 52 maser spots that are assembled into 16 maser features.  
Those maser features were arranged in 11 maser groups, which are labeled  
{\it a} to {\it k}, as shown in 
Figure~\ref{fig:05}. 
The groups {\it a} to {\it e}, and {\it g}, {\it h}, and {\it i} have 
a radial velocity range between $-51.9$ and $-40.5$~km~s$^{-1}$.
The groups {\it f}, {\it j}, and {\it k}, which have been
detected at one epoch, have a radial velocity of
$-52.3$, $-39.3$, and $-37.6$~km~s$^{-1}$, respectively. 
The radial velocities of the identified maser spots are close to the stellar systemic velocity 
({\it V}$_{\mathrm{LSR}}=-$36.16$\pm$0.68~km~s$^{-1}$). 
Groups {\it a} and {\it b} located to the north and southeast of the 
spatial distribution, respectively, are bright and have complicated structures  
while the other groups ({\it c} to {\it k}) have a rather simple structure. 
For the following annual parallax analysis, we selected 16 maser spots among 52 
using the following two criteria. 
(1)~Maser spots were detected with the image S/N greater than or equal to 7.  
(2)~The maser spot was detected at eight epochs or more. 
All the selected maser spots belong to either group {\it a} or {\it b} by chance. 

\subsection{
  Annual Parallax
}\label{sec:04-01}

\begin{figure}
\epsscale{.27}
\plotone{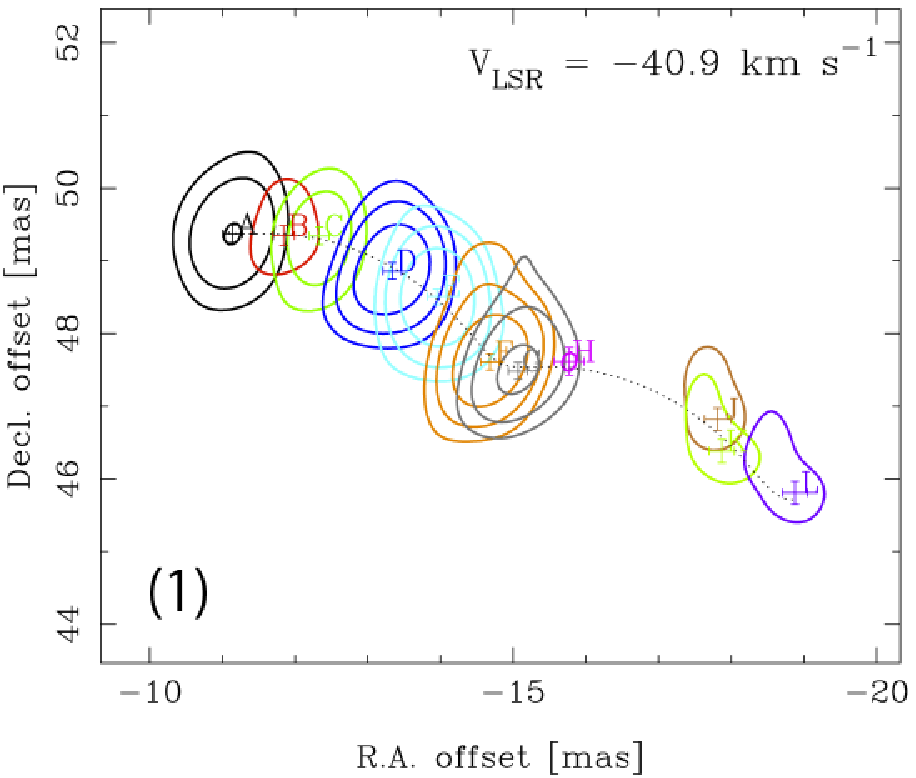}
\plotone{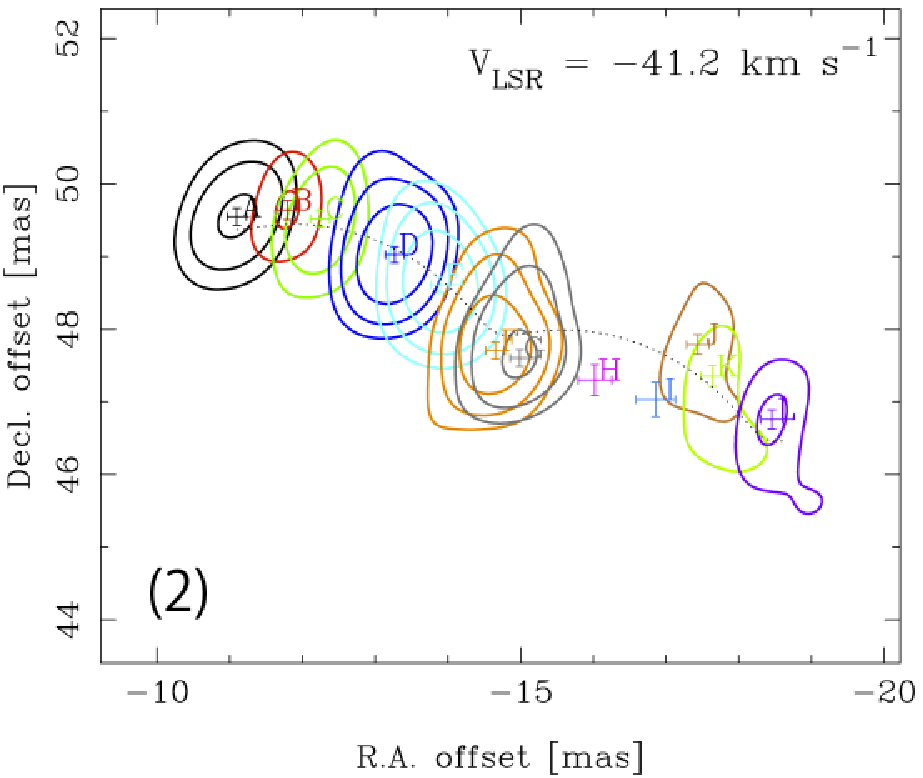}
\plotone{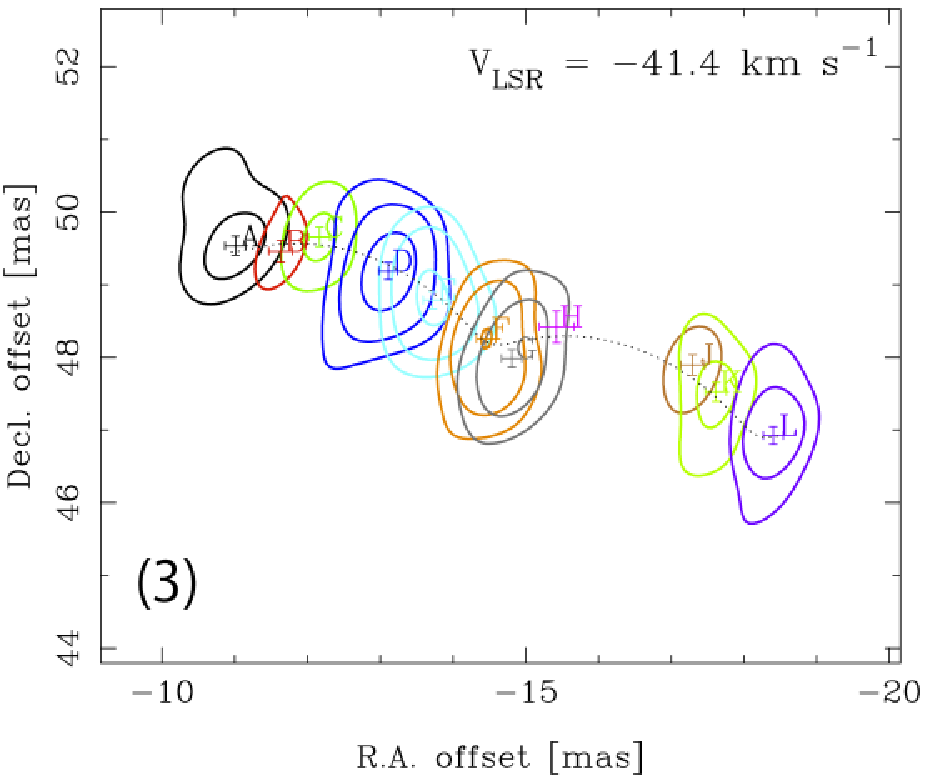}
\plotone{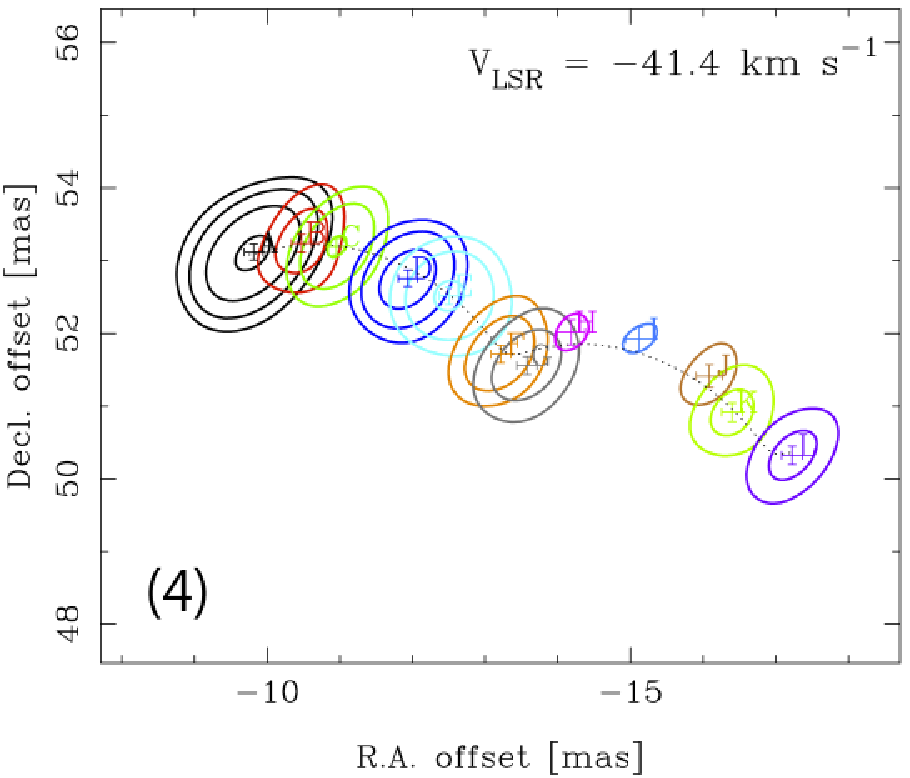}
\plotone{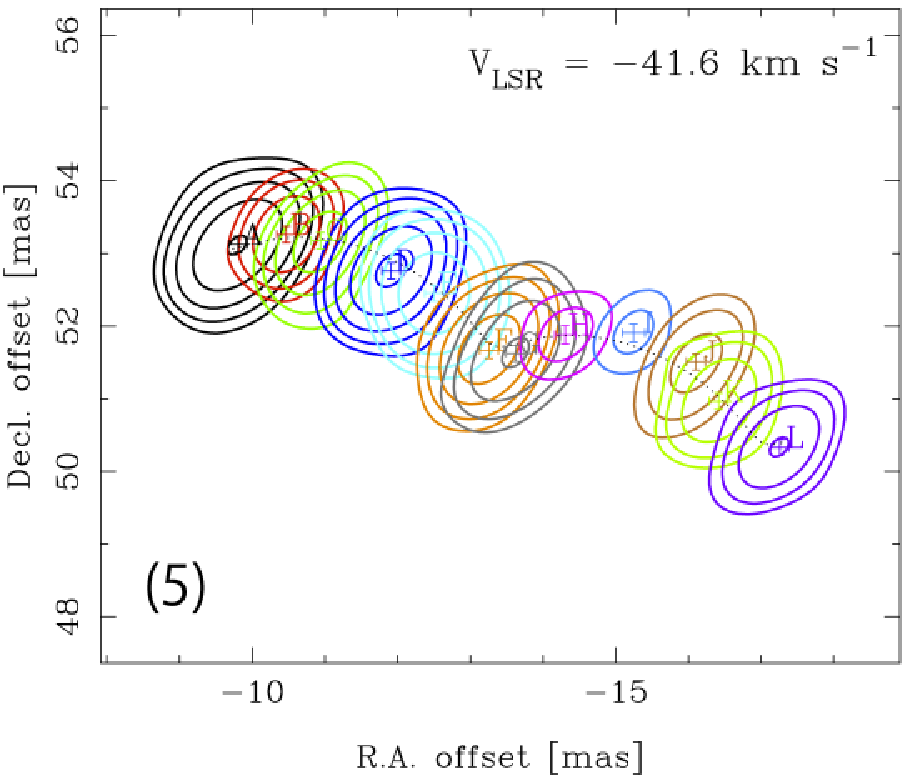}
\plotone{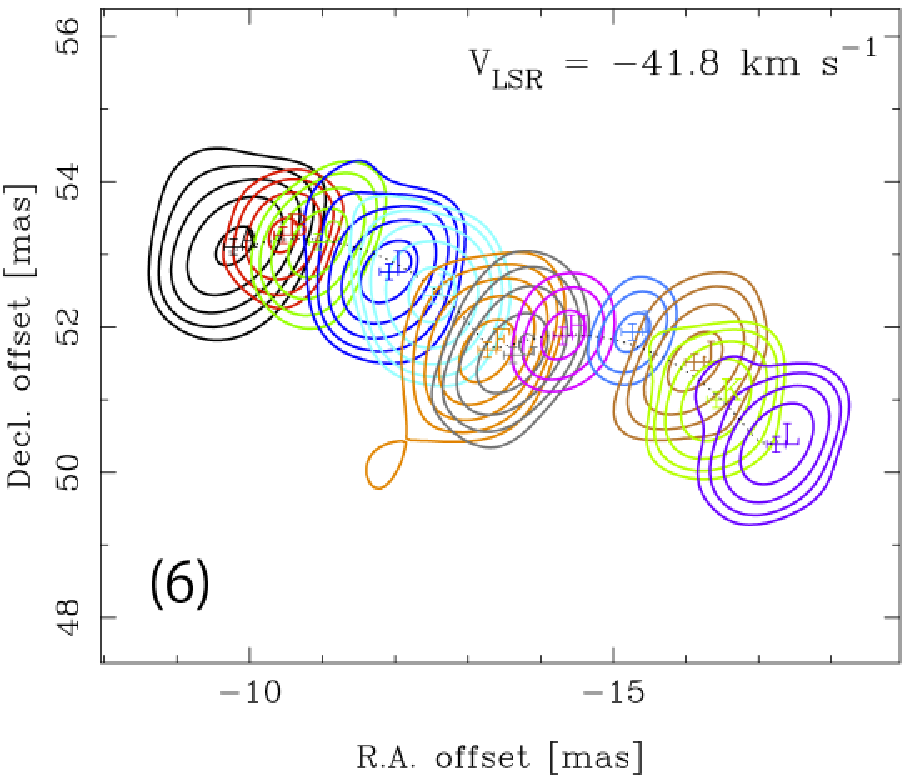}
\plotone{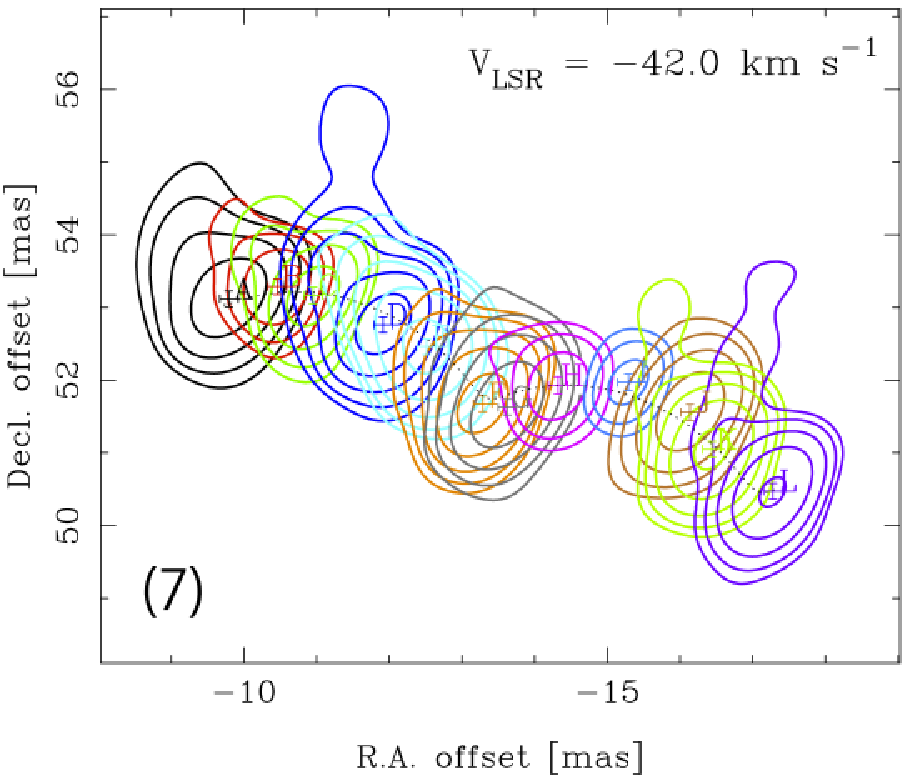}
\plotone{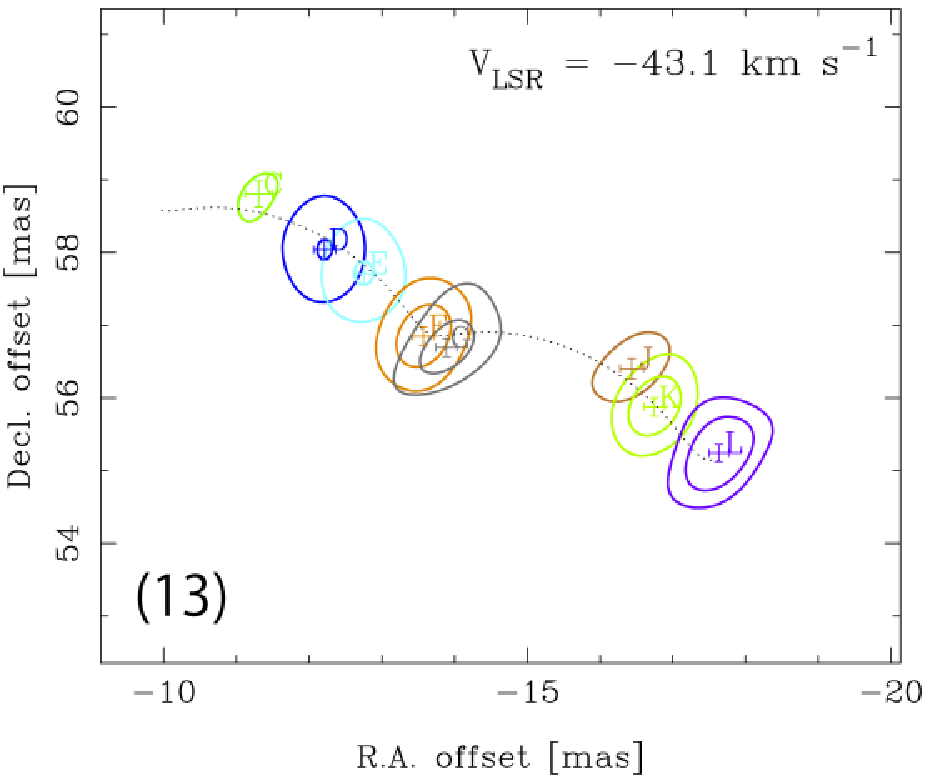}
\plotone{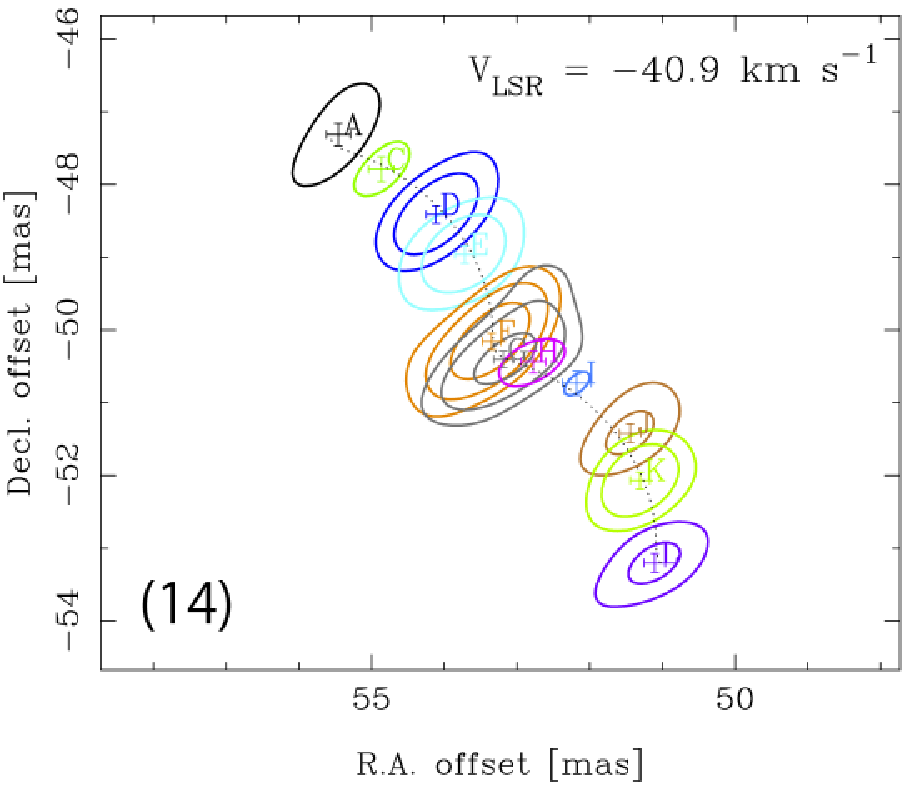}
\plotone{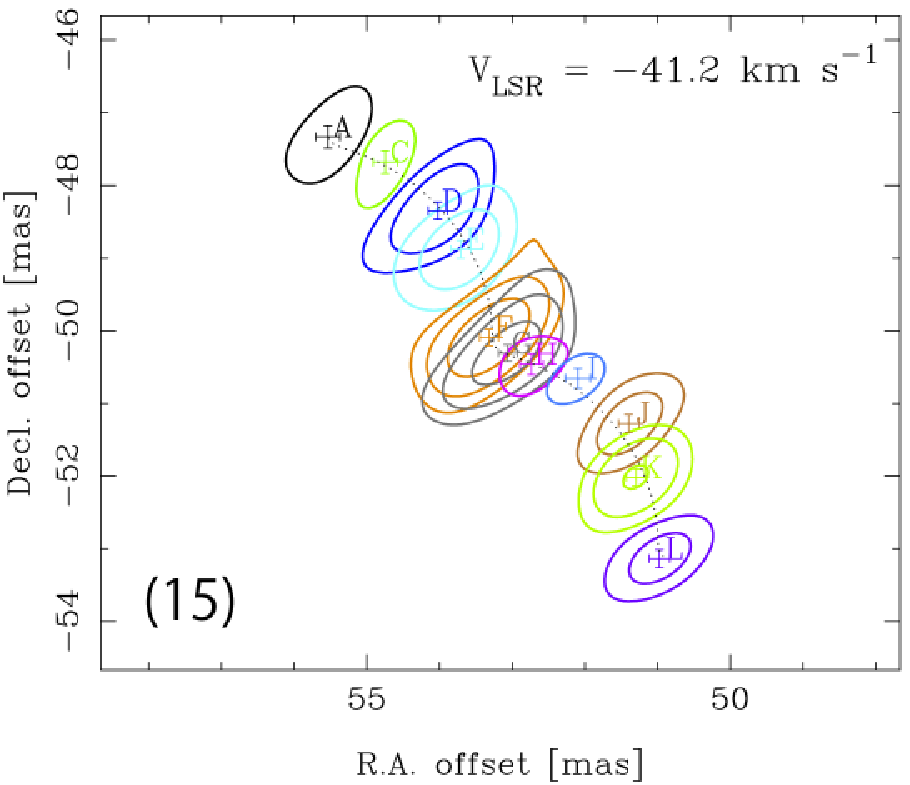}

\caption{
	Spatial motions of 10 isolated maser spots. 
	The color represents an observation epoch as shown in 
	Figure~\ref{fig:05}.
	The dotted lines represent the best-fit annual parallax and proper motion.
	The outermost contours show a 5$\sigma$ noise level increased by a factor of 2.
	The number at the bottom left corner in each of the 
	panels is the maser spot ID as listed in 
	Table~\ref{tbl:05}.
     Positional errors are displayed at the image peak positions. 
	The dashed lines represent the least-squares fitting results of the 
	spatial motion of an annual parallax and a linear motion 
	for the maser peak positions.  
	}
\label{fig:06}
\end{figure}

\begin{figure}
\epsscale{.30}
\plotone{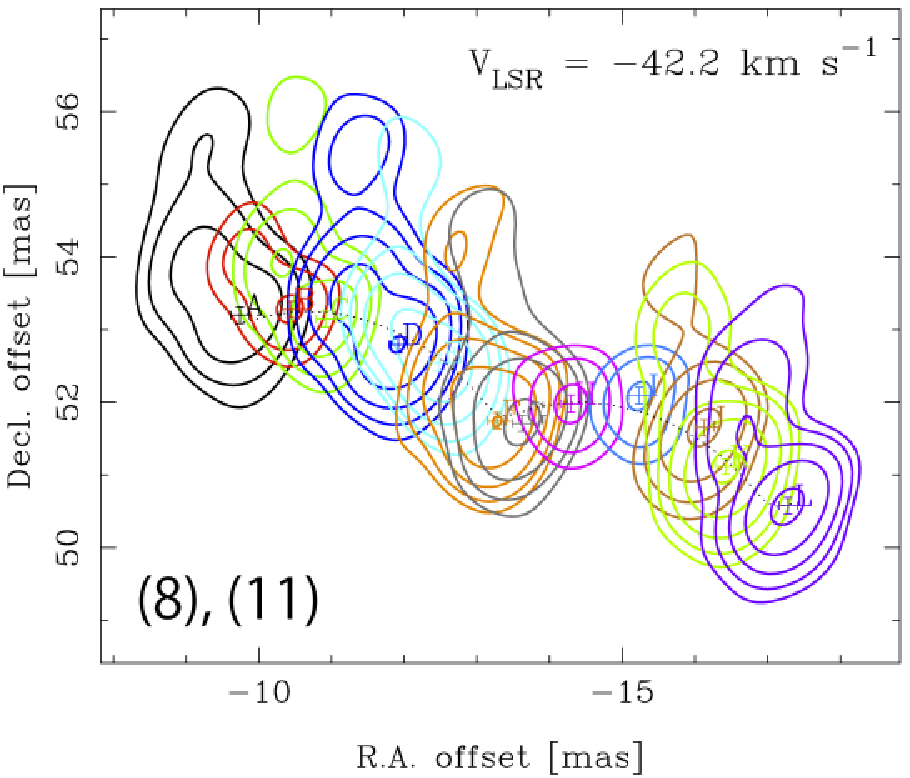}
\plotone{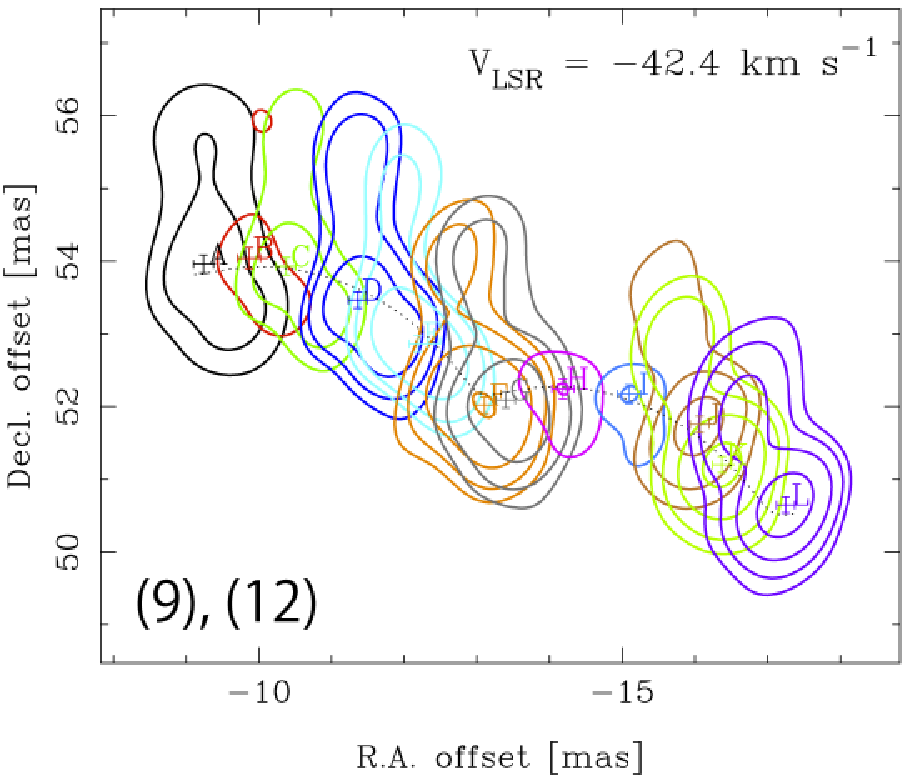}
\plotone{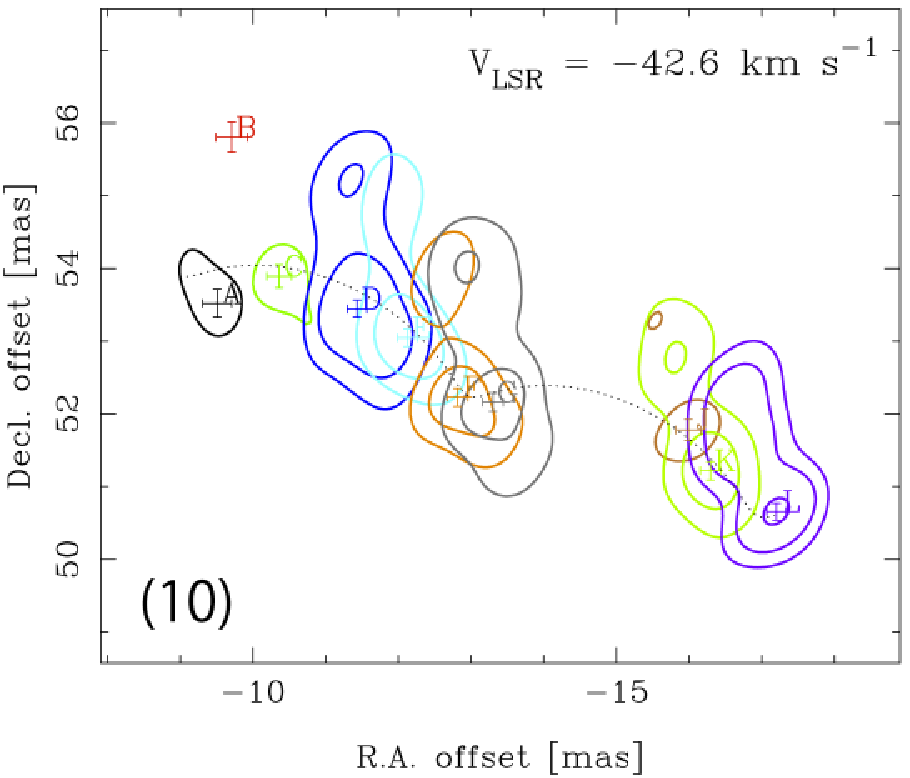}
\plotone{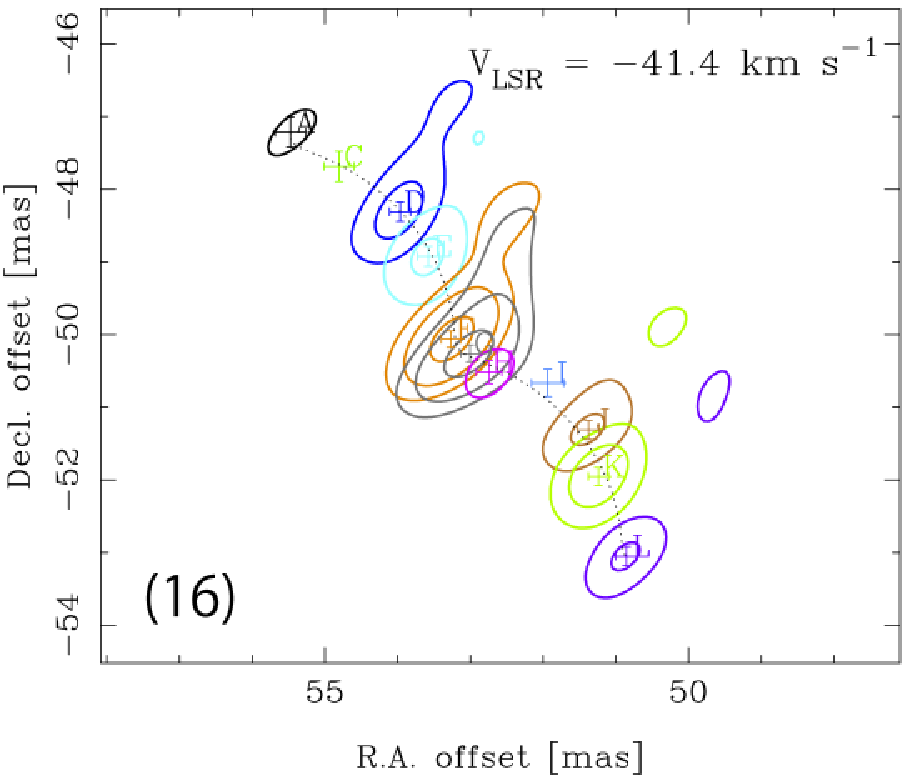}
\caption{
  Same as 
  Figure~\ref{fig:06}, 
  but for the maser spots identified 
  as the blended component. 
  The maser components in the upper two panels have 
  two maser spots while those in the lower panels have a single spot. 
}
\label{fig:07}
\end{figure}

In the first step of the annual parallax analysis we performed a 
Levenberg-Marquardt least-squares model fitting 
of the maser motions in the same way as that 
described by 
\cite{Asaki2010} 
using the peak positions of the selected maser spots. We adopted 
the astrometric error as mentioned in Section~\ref{sec:03-03} and 
a tentative value of a morphology uncertainty 
(a positional uncertainty due to maser morphology variation) 
of $50~\mu$as in the fitting, 
equivalent to 0.1~AU by assuming the distance of 2~kpc to the source. 
Figures~\ref{fig:06} and \ref{fig:07}  
show the images and the fitting results of all the selected maser spots. 
The stellar annual parallax, $\pi^{*}$, was determined by the combined 
fitting with the Levenberg--Marquardt least-squares analysis for all the 
selected spots. The obtained stellar annual parallax was 0.380$\pm$0.011~mas 
at this stage.

\begin{figure}
\epsscale{.70}
\plotone{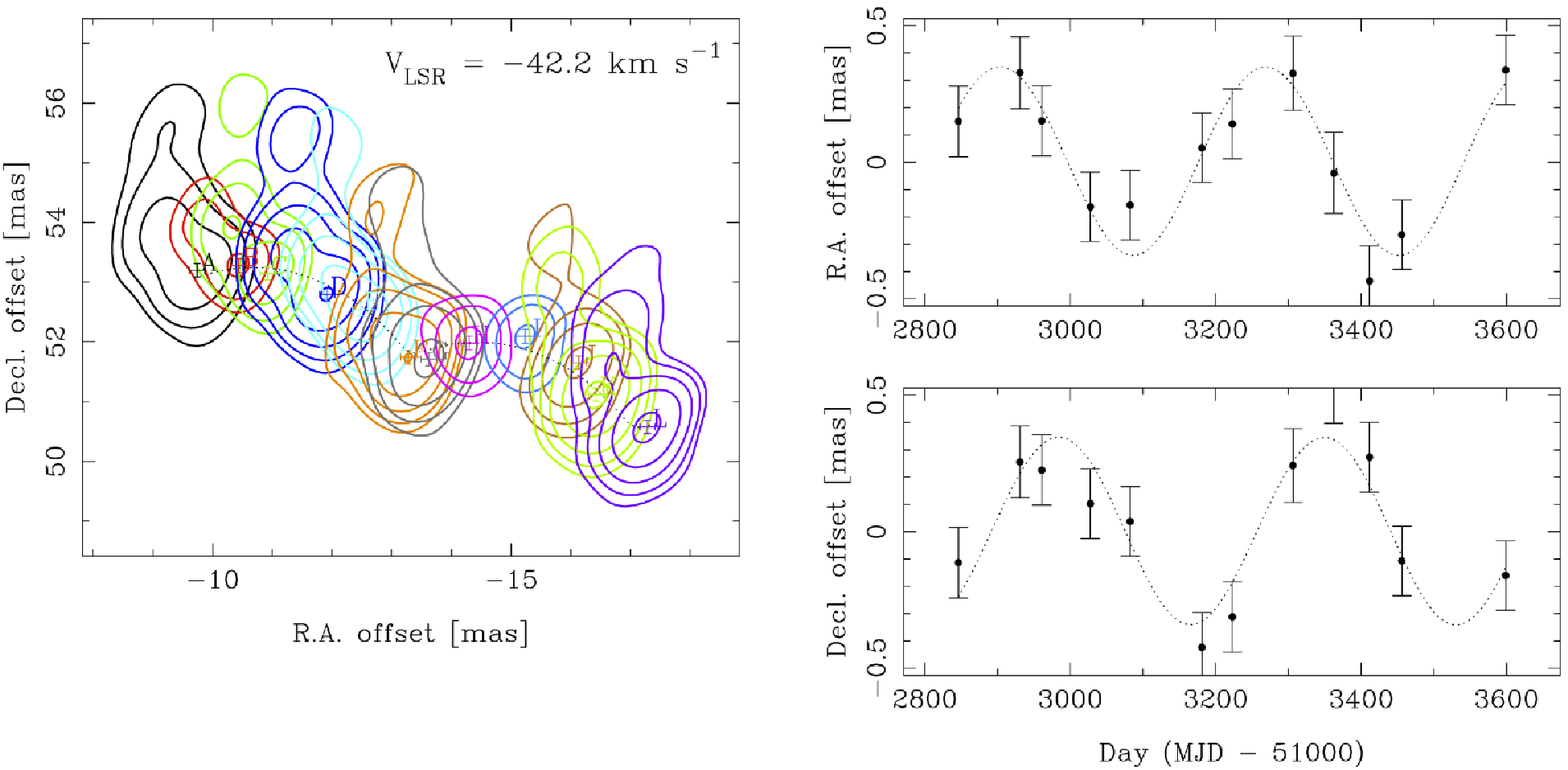}
\plotone{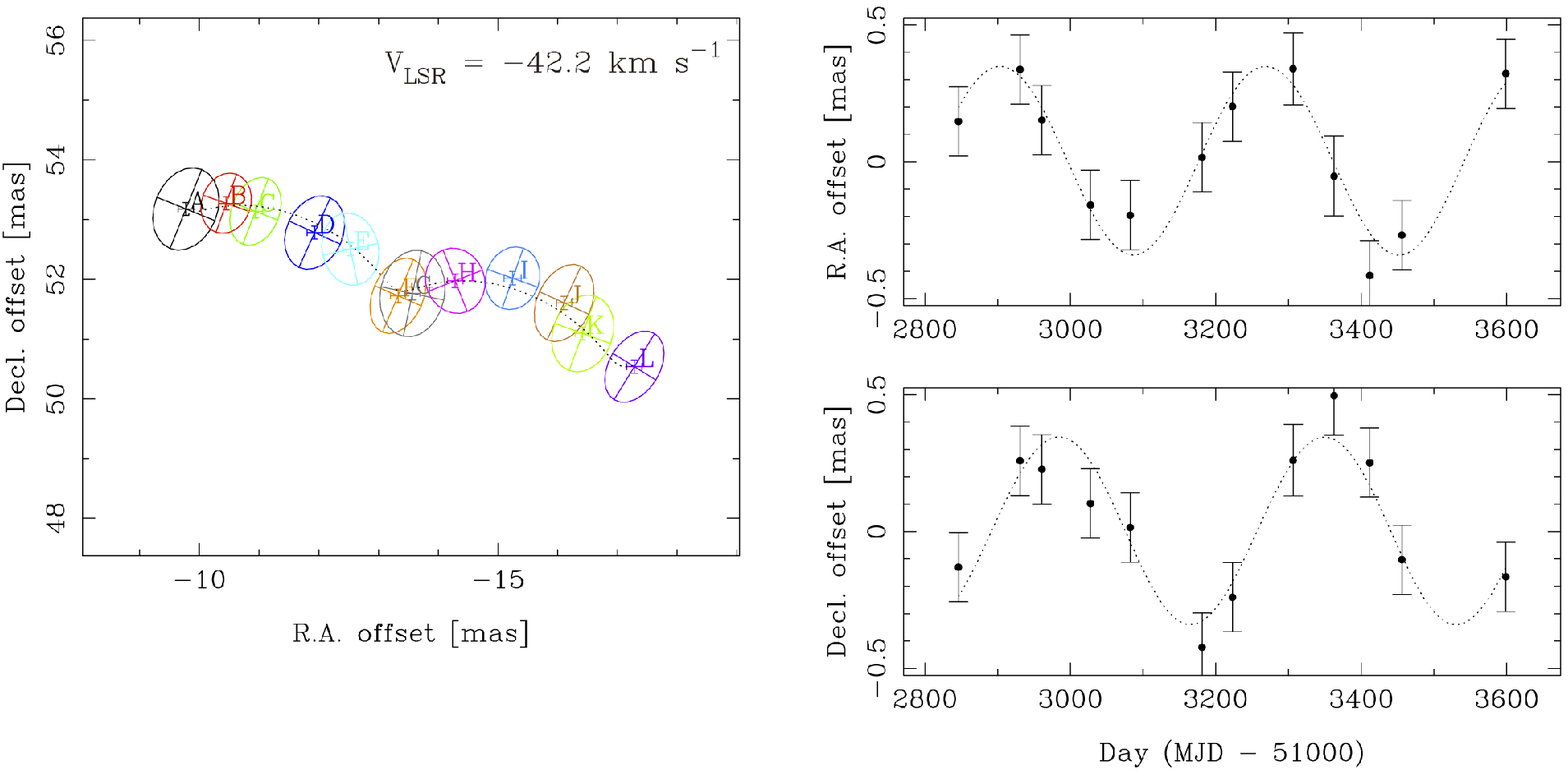}
\caption{
	Least-squares fitting results of the motions of the maser spot with 
	the LSR velocity of $-42.2$~km~s$^{-1}$. 
	Top three panels: least-squares fitting using the image peak positions. 
	Bottom three panels: least-squares fitting using the two-dimensional Gaussian 
	peak positions. 
	The left panels show the maser positions after removing the initial position 
	and the proper motions. 
}
\label{fig:08}
\end{figure}

If we carefully look at 
Figure~\ref{fig:07}, 
the maser spots do not seem to be properly identified with the 
image peak positions because multiple maser spots are spatially blended. 
This is caused when the multiple maser spots, whose size is typically 1~AU 
\citep{Reid1981}, 
are closely located comparing with the synthesized 
beam size of $\sim 1$~mas in this case, equivalent to 2.6~AU at the distance of PZ~Cas. 
To identify an individual maser spot properly, especially for blended 
structures, we performed AIPS JMFIT for the 2D multiple Gaussian 
component fitting of the emissions. Here we refer to a maser spot that can be identified with 
the image peak position as an isolated maser spot, and to a maser spot that cannot 
be identified with a single peak position because of the insufficient spatial resolution as 
a blended component. 
If we can identify 2D Gaussian components using JMFIT, instead of SAD that automatically 
searches multiple spots, in the brightness distribution, we refer to such spots as identified spots. 
Figure~\ref{fig:08} provides an example of the annual parallax least-squares fitting 
for a maser spot using either the image peak position or the 2D Gaussian peak. 
Table~\ref{tbl:04} lists our combined fitting results using the image peak  
only for the isolated spots, only for the blended components, and for all those involved 
in the two groups. We also tried to make another combined fitting using the Gaussian 
peak only for the isolated spots, only for the blended but identified 
spots with JMFIT, and all of them, as listed in the bottom row in  
Table~\ref{tbl:04}. 
The combined fitting results for the isolated spots show little difference between the cases of 
image peak and Gaussian peak. 
However, the fitting result for the blended spots with JMFIT peak positions is 
definitely changed from that with the image 
peak for the blended components and becomes closer to that 
obtained with the isolated spots. If a maser spot is recognized as an isolated spot, 
in other words, if a maser cloud is spatially resolved into multiple spots in the image, 
the image peak position can be treated as the maser position. Because this cannot be 
the case for the maser spots of PZ~Cas, we adopted the Gaussian peak positions in 
our least-squares model fitting. The model fitting results obtained from the individual  
spots are listed in 
Table~\ref{tbl:05}. 
From the combined fitting analysis by using all the selected maser spots, we obtained 
the stellar annual parallax of 0.356~$\pm$~0.011~mas with assumptions of 
the above astrometric error and morphology uncertainty at this stage.  
 
\begin{figure}
\epsscale{.70}
\plotone{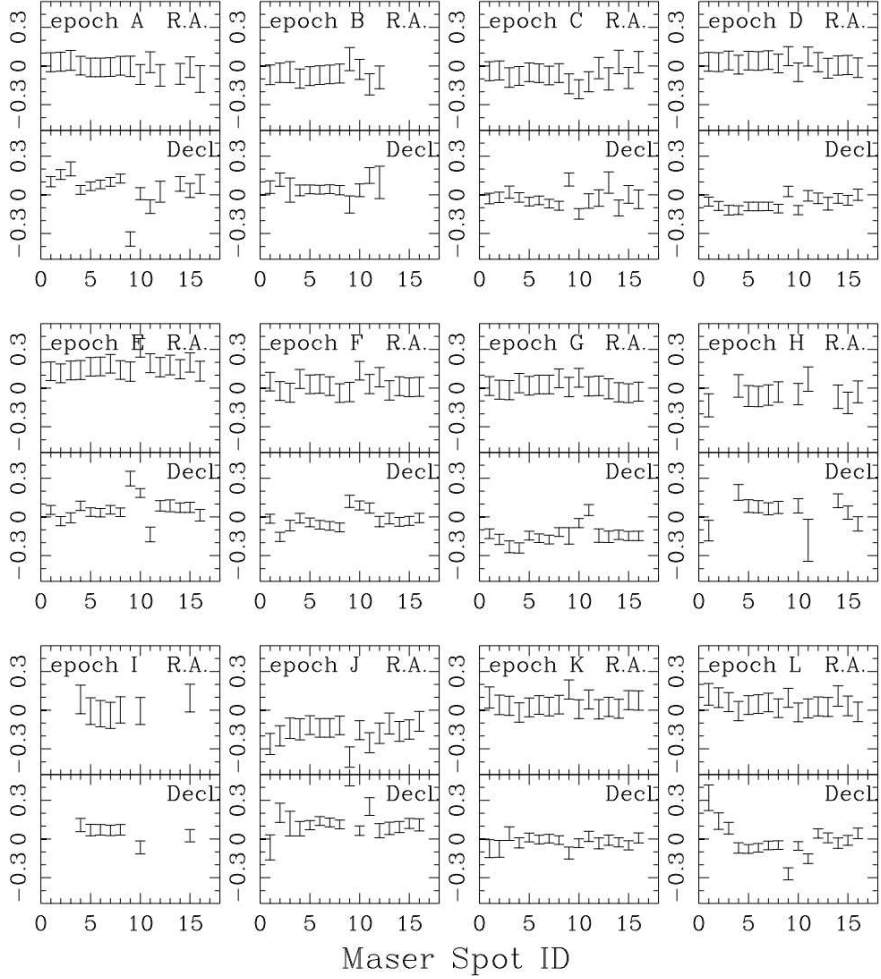}
\caption{
	Position residuals of the 16 maser spots after removing the 
	annual parallax of 0.356 mas, proper motions, and initial 
	positions. The abscissa represents the maser spot ID 
	as shown in Figures~\ref{fig:06} and~\ref{fig:07},
	and the ordinate is the position residual in mas.
	For each epoch, the position residuals in right ascension 
	and declination are shown in the upper and lower panels, 
	respectively. 
}
\label{fig:09}
\end{figure}

Although H$_{2}$O maser features around RSGs are good tracers for astrometry
\citep{Richards2012}, 
the morphology of the maser spots observed with the high-resolution VLBI 
is no longer negligible for the precise maser astrometry. 
These monitoring observations of PZ~Cas have been conducted so often for two 
years that we can investigate the contribution of the morphology uncertainty 
to the maser astrometry. 
Figure~\ref{fig:09} 
shows the position residuals of the 16 selected maser spots after removing the 
stellar annual parallax of 0.356 mas, proper motion, and initial position 
estimated for each of the spots. The standard deviation of the residuals is 
93 and 110~$\mu$as for right ascension and declination, respectively. 
Assuming that those errors have the statistical characteristics of a Gaussian distribution, 
the morphology uncertainty is 59 (~=$\sqrt{93^2-72^2}$~) and 105 (~=$\sqrt{110^2-33^2}$~)~
$\mu$as for right ascension and declination, respectively, 
where 72 and 33 $\mu$as are the astrometric errors discussed in Section 3.3. 
It is unlikely that the morphology uncertainty 
has different values between right ascension and declination, so we adopted 105~$\mu$as 
as the morphology uncertainty to be on the safer side. Based on the trigonometric parallax 
of 0.356~mas, 105~$\mu$as is 0.29~AU at PZ~Cas. We conducted the combined 
fitting with all the maser spots whose positions were determined with JMFIT and 
obtained $\pi^{*}$ of $0.356\pm 0.011$~mas at this stage. 

Note that the above error of the annual parallax is a result of considering the random 
statistical errors from the image S/N and the randomly changing maser spot position, 
as well as a common positional shift to all the maser spots at a specific epoch due to 
the astrometric error.
Because the majority of the maser spots and features in the current parallax measurement
are associated with the maser group~{\it a}, the obtained stellar annual parallax could be 
affected by a time variation of the specific maser features at a certain level. 
Assuming that all the 13 maser spots belonging to group~{\it a} have an identified  
systematic error, the annual parallax error can be evaluated to be 
$0.011\times\sqrt{13}=0.040$~mas. 
However, if we assume that maser spots belonging to a certain maser 
feature have a common systematic error but that some of them have their own 
time variations, the annual parallax error should be estimated to be 
ranging somewhere between 0.011 and 0.040~mas. To evaluate the contributions 
of such a systematic error, we carried out Monte Carlo simulations introduced by 
\cite{Asaki2010} 
for our astrometric analysis. 
First we prepared the imitated data set of the 16 maser spots with position changes 
added by considering the fixed stellar annual parallax and the maser proper motions. 
Second we added an artificial positional offset due to the image S/N, 
the astrometric accuracy, and the morphology uncertainty to each of 
the maser spot positions. The positional offset caused by the image S/N was given to 
each of the maser spots at each epoch independently. 
The positional offset due to the astrometric error was common to all the maser spots 
at a specific epoch but randomly different between the epochs. 
In this paper, we assumed that the positional offset due to the 
morphology uncertainty was common to all the maser spots belonging to a 
specific maser feature at a specific epoch, but randomly different between the 
features and the epochs. 
We assumed the standard deviation of the maser morphology uncertainty 
was 105~$\mu$as, equivalent to the intrinsic size of 0.29~AU. 
In a single simulation trial, we obtained the stellar annual parallax for the imitated data set. 
Thirdly we carried out 200 trials in order to obtain the standard deviation of 
the stellar annual parallax solutions. 
The resultant standard deviation was 0.026~mas which can be treated as the annual 
parallax error in our astrometric analysis. 
The resultant estimate of the stellar annual parallax was 
0.356~$\pm$~0.026~mas, corresponding to $2.81~^{+0.22}_{-0.19}$~kpc.  

\subsection{
  Stellar Position and Proper Motion
}\label{sec:04-02}

\begin{figure}
\epsscale{.65}
\plotone{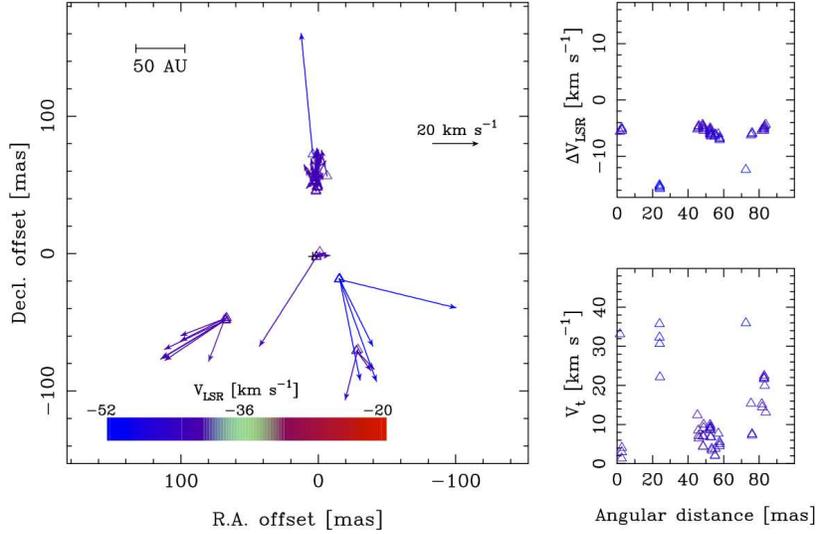}
\caption{
    Internal motions of maser spots around PZ~Cas. 
    The cross bars around the map origin represent the position of PZ~Cas and 
    its positional error obtained with the least-squares fitting. 
    The color represents the radial velocity, 
    and the open circles and triangles are red and blueshifted maser spots with respect to 
    the systemic velocity. 
    Left: spatial distributions of the maser spots with respect to the stellar position 
    (coordinate origin).Top right: radial velocity-position plot. 
    The abscissa is the angular distance from the stellar position, and the ordinate 
    is the radial velocity difference from the 
    systemic velocity. Bottom right: transverse velocity-position plot. 
    }
\label{fig:10}
\end{figure}

Similar to the cases seen in other RSGs 
(VY~CMa,~\citealt{Choi2008}; 
 S~Per,~\citealt{Asaki2010}; 
 VX~Sgr,~\citealt{Kamohara2005};~\citealt{Chen2006}; 
 NML~Cyg,~\citealt{Nagayama2008},~\citealt{Zhang2012b}), 
the maser features around PZ~Cas are widely distributed over a 
$200 \times 200$~square~mas (corresponding to $562 \times 562$~square~AU 
at a distance of 2.81~kpc)
while the velocity of the features ($-52$ to $-37$~km~s$^{-1}$) 
is distributed more closely to the stellar systemic velocity than 
those found in other RSGs are. It is likely that the detected maser groups are 
located at the tangential parts of the shell-like circumstellar envelope, whose 
3D velocity has a small line-of-sight velocity. We determined 
the proper motions of all detected maser spots were identified at least in two epochs. 
Table~\ref{tbl:06} 
lists maser spots whose proper motions and positions were calculated with the 
stellar annual parallax determined above. 
Using the proper motion data listed in 
Tables~\ref{tbl:05} 
and 
\ref{tbl:06}, 
we then conducted the least-squares fitting analysis to obtain the position 
of PZ~Cas and the stellar proper motion based on the spherically expanding 
flow model 
\citep{Imai2011}. 
The obtained stellar proper motion in right ascension and declination, 
$\mu^{*}_\alpha\cos{\delta}=-$3.7 $\pm$ 0.2~mas~yr$^{-1}$ 
and 
$\mu^{*}_\delta=-$2.0 $\pm$ 0.3~mas~yr$^{-1}$, 
respectively. 
The astrometric analysis result for PZ~Cas is listed in 
Table~\ref{tbl:07}. 
The proper motions of the detected maser spots together with the stellar 
position indicated by the cross mark are shown in 
Figure~\ref{fig:10}. 

\section{
  Discussions
}\label{sec:05}

\subsection{
  Distance to PZ Cas and Cas~OB5
}\label{sec:05-01}

We obtained the trigonometric parallax to PZ~Cas, 
$D=2.81^{+0.22}_{-0.19}$~kpc 
as described in Section 4.1. PZ~Cas is thought to be 
a member of the Cas~OB5 association in the 
Perseus spiral arm. 
It is of interest to compare the distance to PZ~Cas determined 
in this paper to distances that have been accepted by other methods. 
\cite{Melnik2009} 
estimated the trigonometric parallax to Cas~OB5 to be $2.0$~kpc from the median 
of trigonometric parallaxes of the member stars obtained with {\it Hipparcos}. 
Since a typical statistical error of {\it Hipparcos} annual parallaxes 
is $\sim 1$~mas, we cannot fully rely on this statistical value for the object. 
Another estimate of the distance to Cas~OB5 is a distance modulus of 12, 
corresponding to the photometric parallax of 2.5~kpc 
\citep{Humphreys1978}. 
VLBI phase-referencing astrometry has revealed that its measured trigonometric parallaxes 
for RSGs are well consistent with photometric parallaxes of the related star clusters within 
$10\%-20$\% 
( S~Per,~\citealt{Asaki2010}; 
 VY~CMa,~\citealt{Zhang2012a}; 
 NML~Cyg,~\citealt{Zhang2012b} ). 
Therefore, it is highly plausible that PZ~Cas is involved in Cas~OB5 from the geometrical 
 viewpoint. 
 
However, the difference between the distances derived from PZ~Cas's annual parallax 
and the photometric parallax of Cas~OB5 is too large to be explained from the viewpoint 
of its extent in depth, compared with the distribution on the sky of the 
member stars.  One of the plausible reasons for the inconsistency is the overestimation of 
$A_{\mathrm{V}}$ to Cas~OB5. 
If we attribute the inconsistency to the $A_{\mathrm{V}}$ value, the difference is 0.16~mag, 
which is comparable to a typical statistical error of $A_{\mathrm{V}}$. 
The {\it Gaia} mission is expected to yield measurements of accurate 
trigonometric distances for much more sampled stars than ever.
 
%
\subsection{
  The Location of PZ~Cas in the H-R Diagram 
}\label{sec:05-02}

\begin{figure}
\epsscale{.55}
\plotone{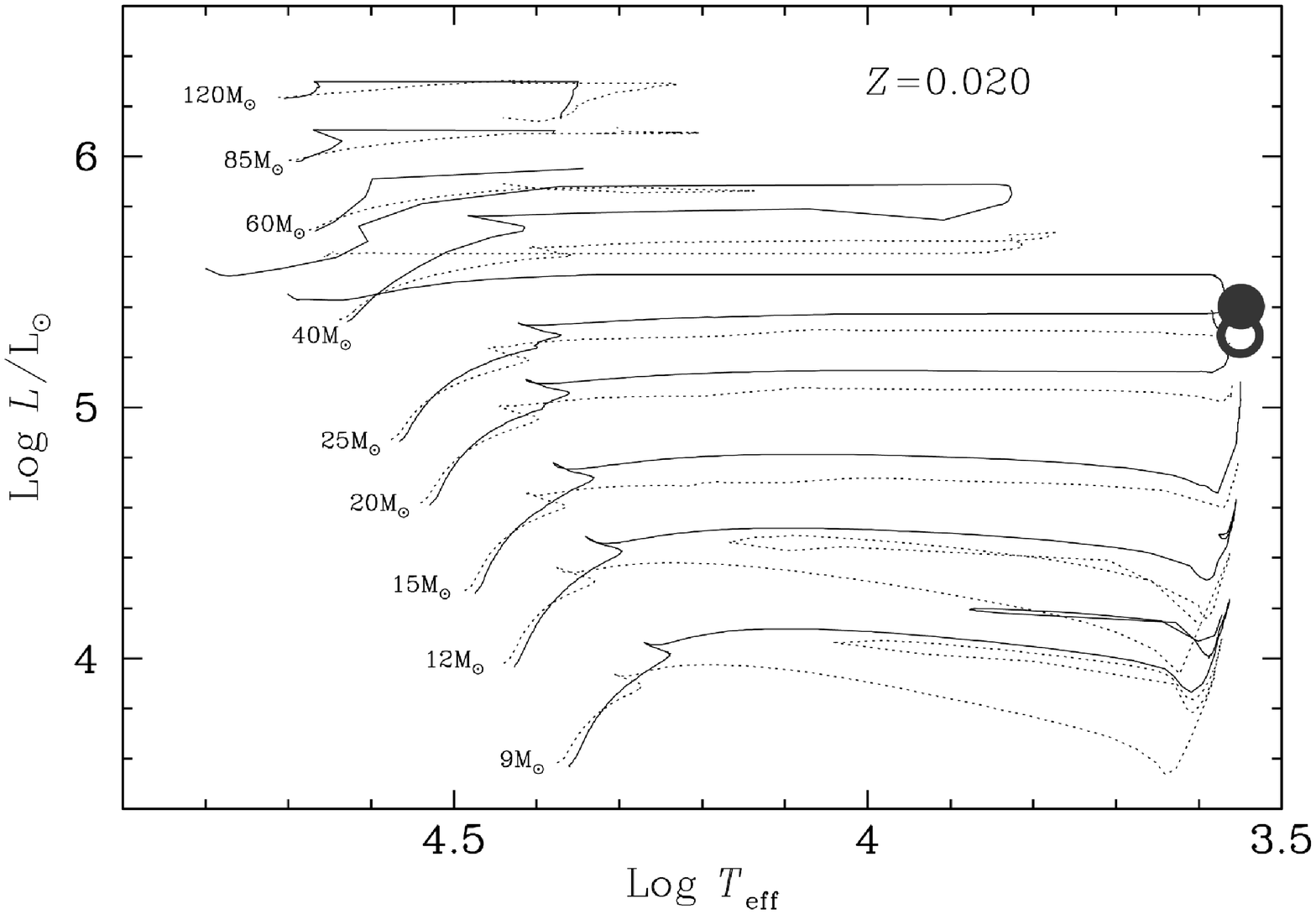}
\caption{
    Locations of PZ~Cas in the H-R diagram 
    \citep{Meynet2003}.
    The open and filled circles represent the locations of PZ~Cas estimated from 
    the distance modulus and our trigonometric parallax, respectively. The effective 
    temperature is assumed to be 3600~K 
    \citep{Levesque2005}. 
}
\label{fig:11}
\end{figure}

The bolometric luminosity, $L=4{\pi}D^{2}F_{\mathrm{bol}}$, 
of PZ~Cas was estimated to be 
$2.0\times10^{5}L_{\odot}$, $2.1\times10^{5}L_{\odot}$, and $1.9\times10^{5}L_{\odot}$ by 
\cite{Josselin2000}, 
\cite{Levesque2005}, and 
\cite{Mauron2011}, 
respectively, with the distance modulus of Cas~OB5 of 12 
\citep{Humphreys1978}. 
To derive the mass and age of PZ~Cas, we first rescaled the luminosity values 
with our trigonometric parallax to 
$2.5\times10^{5}L_{\odot}$, $2.7\times10^{5}L_{\odot}$, and $2.4 \times 10^{5}L_{\odot}$, 
respectively. 
\cite{Levesque2005} 
estimated the effective temperature of PZ~Cas to be 3600~K from the MARCS
stellar atmosphere models and the absolute spectro-photometry. 
Using this effective temperature, we ploted PZ~Cas in the H-R diagram reported by 
\cite{Meynet2003}, 
as shown in 
Figure~\ref{fig:11}. 
In this figure 
the filled circles represent the re-estimated luminosities based on 
\cite{Josselin2000}, 
\cite{Levesque2005}, and 
\cite{Mauron2011}, 
while the open circles represent the originally reported ones. 
The luminosity previously estimated by using the distance modulus 
yields the initial mass of around 20~$M_{\odot}$ while,  
using our trigonometric parallax, it is estimated to be 
$\sim$25~$M_{\odot}$. 

According to evolutionary tracks shown in Figures~1 and 2 of 
\cite{Meynet2003}, 
the lifetime is estimated to be $\sim 8$~Myr for an initial mass of 
25~$M_{\odot}$ and $\sim 10$~Myr for an initial mass of 20~$M_{\odot}$. 
Since the ages of 
four open clusters in Cas~OB5 have been reported to be 
$120\pm20$~Myr for NGC~7790 
\citep{Gupta2000},  
and 30.2, 45.7, and 13.2~Myr for 
NGC~7788,  Frolov~1, and King~12, respectively  
\citep{Kharchenko2005}, 
it is expected that PZ~Cas was born after the 
first generation in Cas~OB5. 

%
\subsection{
  Three~Dimensional Motion of PZ~Cas in the Milky Way 
}\label{sec:05-03}

The stellar proper motion and the radial velocity of PZ~Cas enabled us to discuss its 3D motion 
in the Milky Way. In the analysis, we adopted the Galactocentric distance to the Sun of 8.5~kpc, 
and the flat Galactic rotation curve with the rotation velocity of 220~km~s$^{-1}$. 
For the solar motion, we adopted 
the latest value obtained by 
\cite{Schonrich2010}, 
$11.10\pm0.75$~km~s$^{-1}$ toward the direction of the Galactic center ($U_{\odot}$ component), 
$12.24\pm0.47$~km~s$^{-1}$ toward the direction of the Galactic rotation ($V_{\odot}$ component), 
and 
$7.25\pm0.37$~km~s$^{-1}$ toward the north pole ($W_{\odot}$ component). 
Since H$_{2}$O maser features around RSGs are spherically distributed,  
it is hardly plausible that an uncertainty in the stellar proper motion was caused by 
the anisotropic distribution, as those in the case of outflows of star forming regions 
has to be considered 
\citep[e.g.,][]{Sanna2009}.   
Therefore, we did not introduce an additional uncertainty 
of the PZ~Cas stellar motion 
other than the statistical error described in Section~\ref{sec:04-02}  
in this error estimation. 
We followed the methods described by 
\cite{Johnson1987} 
and 
\cite{Reid2009} 
to derive the peculiar motion and its error. 
The resultant peculiar motion of PZ~Cas was 
($U_{\mathrm{s}}$,~$V_{\mathrm{s}}$,~$W_{\mathrm{s}}$)$=$
($22.8 \pm 1.5$,~$7.1 \pm 4.4$,~$-5.7 \pm 4.4$)~km~s$^{-1}$. 

\begin{figure}
\epsscale{1.0}
\includegraphics[width=10cm]{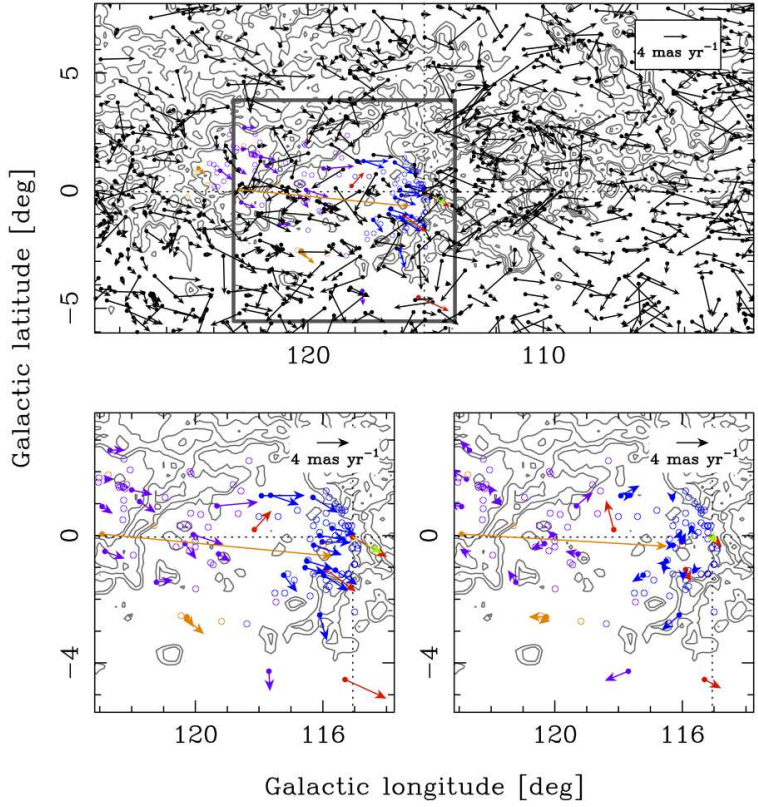}
\caption{ 
    Top: proper motions of the Cas~OB5, Cas~OB4, and Cas~OB7 members 
    as well as stars with the annual parallaxes less than 5~mas and 
    the proper motions less than 8~mas~yr$^{-1}$ in the {\it Hipparcos} catalog. 
    The blue and red circles represent OB stars and RSGs, respectively, 
    which have been identified as Cas~OB5 members. 
    The same goes for the Cas~OB4 and Cas~OB7. The purple and orange circles represent OB stars 
    and RSGs, respectively.
    The filled and open circles denote the member stars found and not found in the 
    {\it Hipparcos} catalog, respectively. 
    The black points denote the stars whose proper motions are reported in the 
    {\it Hipparcos} catalog but not identified with Cas~OB5, Cas~OB4, and Cas~OB7 members. 
    The arrows represent proper motions listed in the {\it Hipparcos} catalog. 
    The position of PZ~Cas is located at the crossing point of the horizontal and vertical 
    dotted lines. The green arrow represents the proper motion reported in this paper. 
    The background contour is the CO map reported by 
    \cite{Dame2001}.
    Bottom left: an enlarged picture of the area surrounded by the dashed square in the top panel. 
    Bottom right: same as the bottom left panel, but the average proper motion of the Cas~OB5 members 
    was subtracted from each proper motion. 
}
\label{fig:12}
\end{figure}

\begin{figure}
\epsscale{1.0}
\includegraphics[width=120mm]{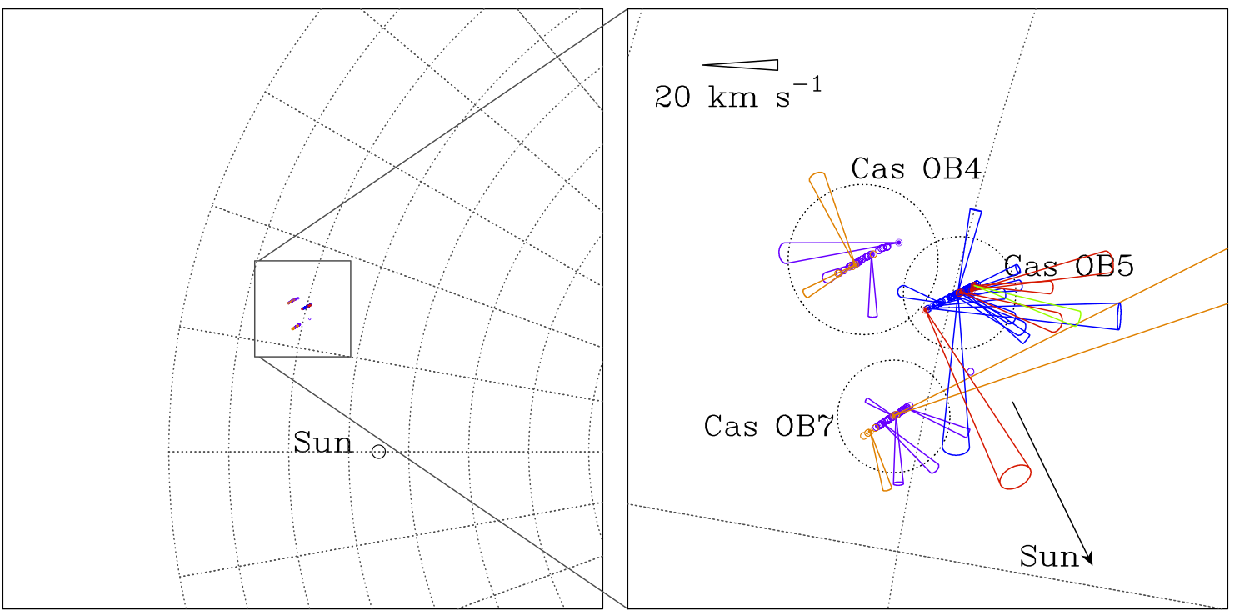}\\
\includegraphics[width=120mm]{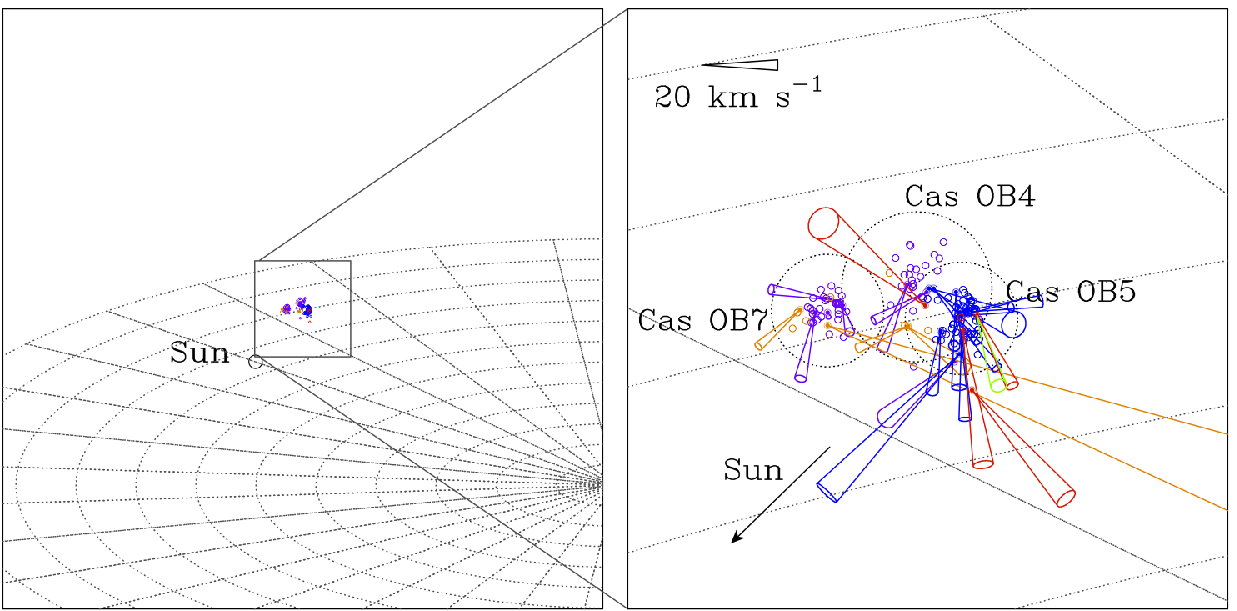}
\caption{ 
    Three-dimensional views of the positions and peculiar motions of the member 
    stars of Cas~OB5 
    as well as Cas~OB4 and Cas~OB7. Note that the distance to Cas~OB5 
    is assumed to be 2.8~kpc (this paper) while those of Cas~OB4 and Cas~OB7 
    are assumed to be 2.9 and 2.5~kpc, respectively, determined from the photometric 
    parallaxes  
\citep{Humphreys1978}. 
    The blue and red points are OB stars and RSGs of Cas~OB5, respectively, 
    listed in 
    Table~\ref{tbl:08}. 
    The green point represents PZ~Cas's peculiar motion obtained in this paper. 
    The purple and orange points are OB stars and RSGs of the other two OB associations, 
    respectively, listed in 
    Table~\ref{tbl:09}. 
    Conicals in the enlarged right plots represent the peculiar 
    motions of the member stars after removing the flat Galactic rotation curve 
    of 220~km~s$^{-1}$. A thick gray dashed line represents the position of the 
    Perseus spiral arm   
    \citep{Asaki2010}. 
}
\label{fig:13}
\end{figure}

Different from peculiar motions of Galactic high-mass star-forming regions (HMSFRs) 
typically showing negative 
$V_{\mathrm{s}}$ 
components ($-10$ to $-20$~km~s$^{-1}$; e.g., 
\citealt{Reid2009}), 
PZ~Cas has a large peculiar motion not in the $V_{\mathrm{s}}$ component but in $U_{\mathrm{s}}$. 
At first, in order to discuss whether PZ~Cas's peculiar motion is due to the internal motion inside 
Cas~OB5, we investigated proper motions of the  
member stars of Cas~OB5 using the {\it Hipparcos} catalog 
\citep{Hog2000}. 
Positions and velocities of 
Cas~OB5 member stars (OB stars and RSGs) cataloged in 
\cite{Garmany1992} 
with compiled information from previous literatures 
\citep{
Evans1967,
Famaey2005,
Fehrenbach1996,
Humphreys1970, 
Valdes2004,
Leeuwen2007, 
Wilson1953}
are listed in 
Table~\ref{tbl:08}. 
Figure~\ref{fig:12} 
shows the positions and proper motions of the member stars listed in  
Table~\ref{tbl:08} 
as well as stars with 
annual parallax $< 5$~mas and 
proper motion $< 8$~mas~yr$^{-1}$ 
in order to roughly exclude foreground stars. In 
Figure~\ref{fig:12}, 
the $J=1\rightarrow 0$ CO contour map of a Galactic plane survey 
\citep{Dame2001} 
taken from {\it SkyView} (Internet's Virtual Telescope)
\footnote{
SkyView web: http://skyview.gsfc.nasa.gov/
} is superimposed. 
We obtained the unweighted average of those proper motions 
($\mu_\alpha\cos{\delta}=-2.8\pm1.2$ and $\mu_\delta =-1.9\pm0.8$~mas~yr$^{-1}$)  
which is well consistent with 
Cas~OB5's peculiar motion obtained by 
\cite{Melnik2009}
($\mu_\alpha\cos{\delta}=-3.8 \pm 1.2$ and $\mu_\delta =-1.4 \pm 0.8$~ mas yr$^{-1}$). 
Figure~\ref{fig:13} 
shows 3D views of the relative positions and peculiar motions of PZ~Cas and 
the member stars of Cas~OB5 whose radial velocities can be found in previous 
literatures as listed in 
Table~\ref{tbl:08}. 
Here we adopted PZ~Cas's trigonometric distance for Cas~OB5. 
The averaged peculiar motion of those member stars was  
($\bar{U}$,~$\bar{V}$,~$\bar{W}$)$=$($8 \pm 14$,~$-1 \pm 18$,~$-9 \pm 18$)~km~s$^{-1}$ 
whose $U$ component was prominent comparing with the other two.  
Therefore, we concluded that the peculiar motion of PZ~Cas with 
the large $U_{\mathrm{s}}$ component reflects Cas~OB5's peculiar motion. 

Why does PZ~Cas, or Cas~OB5, have such a large peculiar motion in the $U$ component?
Since the member stars including PZ~Cas are located on the western side of 
a large shell-like structure molecular gas (H$_{\mathrm{I}}$ super shell) with the center 
position of a Galactic coordinate of ($118^{\circ}$,~$-1^{\circ}$),
one may consider that 
supernovae in precedent giant molecular clouds triggered the formation of those 
stars and that they have an expanding motion
\citep{Sato2008}. 
We investigated {\it Hipparcos} proper motions of the member stars of Cas~OB5, 
as well as those of Cas~OB4 and Cas~OB7 located on the eastern side of the 
shell-like structure, as shown in 
Figure~\ref{fig:12}. 
We also investigated 3D relative positions and peculiar motions of those of Cas~OB4 
and Cas~OB7, as listed in 
Table~\ref{tbl:09}.
We adopted the photometric parallaxes for Cas~OB4 and Cas~OB7 reported by 
\cite{Garmany1992}. 
Figure~\ref{fig:13} 
shows 3D views of the positions and proper motions of the memeber stars of Cas~OB5 
as well as Cas~OB4 and Cas~OB7.
We could hardly find clear evidence 
of an expanding motion of the member stars, either in the proper motions or 
the 3D motions between those three OB associations.
One of the member stars of Cas~OB7, HD~4842, has a much larger proper 
motion than those of the others, even if the {\it Hipparcos} trigonometric parallax of 
$1.46 \pm 1.21$~mas is considered. Therefore, this star could be a runaway star 
\citep{Fujii2011}.

VLBI astrometric observations of Galactic maser sources on Perseus spiral 
arm have revealed large peculiar motions of Galactic HMSFRs. 
According to 
 \cite{Sakai2012}, 
 33 sources whose peculiar motions have been measured in the Perseus and Outer arms 
 show that these peculiar motions are almost located in the fourth quadrant in the $U-V$ plane
 ($U~>$~0 and $V~<$~0).
 The peculiar motion of Cas~OB5 is located in the fourth quadrant while that of PZ~Cas is located 
 in the first quadrant. 
Such a peculiar motion of the OB association may indicate the standard density wave theory 
\citep{Lin1964}. 
However, it is hard to explain the velocity dispersion of the peculiar motions of the HMSFRs 
from the density wave theory.

Such a large velocity dispersion in peculiar motions can be seen for the Galactic OB 
associations within 3~kpc from the Sun 
\citep{Melnik2009}. 
The peculiar motion of Cas~OB5 hits upon  an interesting idea of 
``time-dependent potential spiral arms''. 
\cite{Baba2009} 
reported on results of large $N$-body and smoothed particle hydrodynamics (SPH) simulations 
that calculated the motions of multi-phase, self-gravity gas particles in self-induced stellar spiral 
arms in a galactic disk. In their simulations, spiral arms are developed by the self-gravity but 
cannot keep their forms even in one galactic rotation evolution. 
Other simulation results by 
\cite{Wada2011} 
suggest that  a massive cloud falls into a spiral arm's potential from both sides, 
that is, there are flows 
of cold gas with various velocities. The flows converge into condensations of cold gas 
near the bottom of the potential well. The relative velocities of the gas to the stellar 
spirals are comparable to the random motion of the gas. 
Another interesting point reported by 
\cite{Wada2011} 
is that supernovae feedback is not essential to change the velocity field of the 
gas near the spiral potential. 
Each of the cold massive clouds and stars with ages younger than 30~Myr 
may have a peculiar motion with a velocity dispersion of $\sim 30$~km~s$^{-1}$, 
which is not predicted by the density wave theory.  
Although we have not had samples large enough to discuss the whole structure 
of the Milky Way, 
the self-consistent gravity spiral galaxy model can explain 
those peculiar motions of HMSFRs in the Milky Way.
In other words, a live spiral structure model can fit to the recent sample of the HMSFRs' 
spatial motions including Cas~OB5. 
A deep understanding of the dynamics of the Galactic spiral arms and the origin of 
the peculiar motions of HMSFRs will be possible in the near future when massive numerical 
simulations can be combined with much larger VLBI astrometric samples 
which only can be conducted toward the Galactic plane obscured by dense gas.

\section{
  Conclusions
}\label{sec:06}

We have conducted phase-referencing VLBI monitoring observations for 
the H$_{2}$O masers of the Galactic RSG, PZ~Cas. The distance to PZ~Cas 
was determined to be $2.81^{+0.22}_{-0.19}$~kpc from the derived annual 
parallax in the maser spot motions. We estimated the initial mass of PZ~Cas as 
$\sim 25$~$M_{\odot}$ using the re-scaled stellar luminosity with 
the annual parallax. 

Together with the stellar radial velocity and obtained proper motion, we investigated 
the 3D motion of PZ~Cas in the Milky Way, 
showing the non-circular motion with a rather large velocity toward the Galactic 
center. We compared the proper motion and 3D peculiar motion of PZ~Cas with those 
of the member stars of Cas~OB5 using the proper motions cataloged in 
{\it Hipparcos} and their radial velocities. We found that the motion of PZ~Cas in 
the Milky Way is close to that of the OB association. This peculiar motion is not 
aligned with those of the Galactic HMSFRs. 
The observed peculiar motion supports a live spiral arm model recently revealed 
by large $N$-body/SPH numerical simulations. 


\acknowledgments

The VERA/Mizusawa VLBI observatory is a branch of the National Astronomical Observatory 
of Japan, National Institutes of Nature Sciences. 
The authors acknowledge the use of NASA's SkyView facility 
(http://skyview.gsfc.nasa.gov) located at NASA Goddard Space Flight Center. 
K.~Kusuno was financially supported by the Center for the Promotion of Integrated 
Science of the Graduate University of Advanced Studies for this publication. 

\newpage

\clearpage
\clearpage
%
%
\clearpage
%
%
\clearpage
%
%
\clearpage
%
%
\clearpage
%
%
%
%
\clearpage
%
%
\clearpage
%
%
\clearpage
%
%
%
\clearpage
%
%
\clearpage
%
%
\clearpage
%
%
%
%
\clearpage
%
%

%
%
\begin{table}
  \begin{center}
  \caption{Observing Epochs of PZ~Cas Astrometric Monitoring Observations}
  \label{tbl:01}
  \begin{tabular}{clcp{70mm}}
\tableline\tableline
     Epoch & 
     \multicolumn{1}{c}{Date} &  Time Range \\
                  &          &  (UTC)  \\
\tableline
A   & 2006 Apr 20  & 19:41-05:20 \\
B   & 2006 Jul 14   & 14:21-23:10 \\
C   & 2006 Aug 13 & 12:21-21:35 \\
D   & 2006 Oct 19   &  08:01-17:20 \\  
E    & 2006 Dec 13  & 04:21-13:40 \\ 
F    & 2007 Mar 22   & 00:21-08:17 \\  
G\tablenotemark{a}    & 2007 May 2     & 18:51-04:40 \\  
H\tablenotemark{b}    & 2007 Jul 25     & 13:20-23:04 \\ 
I \tablenotemark{c}   & 2007 Sep 19     & 10:20-20:10 \\  
J    & 2007 Nov 7      & 07:20-17:10 \\
K    & 2007 Dec 22   & 04:25-14:15 \\  
L    & 2008  May 12   & 19:00-04:50 \\  
\tableline
\tableline
  \end{tabular}
  \tablenotetext{a}{Signal-to-noise ratio for all the baselines was unexpectedly low. }
  \tablenotetext{b}{$T_{\mathrm{sys}}$ data of ISHIGAKI was used in place of OGASA20. }
  \tablenotetext{c}{ IRIKI did not attend the observation. }
\end{center}
\end{table}
\clearpage
%
%
\begin{table}
  \begin{center}
  \caption{Frequency Allocation for PZ~Cas and J2339+6010 at Epoch~F }
  \label{tbl:02}
  \begin{tabular}{c | l | l}
\tableline\tableline
Beam      & \multicolumn{1}{c|}{A}  &  \multicolumn{1}{c}{B}   \\
\tableline
Source   & \multicolumn{1}{c|}{PZ~Cas} & \multicolumn{1}{c}{J2339+6010}     \\
\tableline
Base band number & \multicolumn{1}{c|}{1} & \multicolumn{1}{c}{14} \\
\tableline
&    & 22.106-22.122                    \\
&    & 22.122-22.138                    \\
&    & 22.138-22.154                     \\
&    & 22.154-22.170                  \\
&    & 22.170-22.186                   \\
Frequency range &    & 22.186-22.202                  \\
\ (GHz) & 22.234-22.242   & 22.202-22.218      \\
&                                   & 22.234-22.250               \\
        &                                   & 22.250-22.266     \\
        &                                   & 22.266-22.282        \\
                    &                                   & 22.282-22.298        \\
&    & 22.298-22.314             \\
&    & 22.314-22.330             \\
&    & 22.330-22.346               \\ 
\tableline
Frequency spacing    & 15.625~kHz\tablenotemark{a} (512 channels)  & 250.0~kHz (64 channels)    \\
\tableline
\tableline
  \end{tabular}
\tablenotetext{a}{
     The velocity spacing of the A beam for the H$_{2}$O masers is 0.2107~km~s$^{-1}$. }
\end{center}
\end{table}
\clearpage
%
%
\begin{table}
  \begin{center}
  \caption{Phase Tracking Center Positions of the Observed Sources }
  \label{tbl:03}
  \begin{tabular}{lcc}
\tableline\tableline
    
  & Right Ascension (J2000)
  & Declination (J2000) \\
  \tableline
    PZ~Cas
  & $23^{\mathrm{h}} 44^{\mathrm{m}} 03^{\mathrm{s}}.281900$
  & $+61^{\circ} 47' 22''.182000$ \\
    J2339+6010
  & $23^{\mathrm{h}} 39^{\mathrm{m}} 21^{\mathrm{s}}.125210$
  & $+60^{\circ} 10' 11''.849000$ \\
\tableline\tableline
  \end{tabular}
\end{center}
\end{table}
\clearpage
%
%
%
\begin{table}
  \begin{center}
  \caption{Annual Parallax Results (mas) of PZ~Cas }
  \label{tbl:04}
  \begin{tabular}{cccc}
\tableline\tableline
  Spot Identification Method
   & Blended Component
   & Isolated Component
   & All \\      
\tableline           
   Image peak
  & 0.402 $\pm$ 0.021  
  & 0.371 $\pm$ 0.013
  & 0.380 $\pm$ 0.011 \\
\tableline
   2-D Gaussian peak  
  & 0.352 $\pm$ 0.017
  & 0.358 $\pm$ 0.013
  & 0.356 $\pm$ 0.011 \\
\tableline\tableline
  \end{tabular}
\end{center}
\end{table}
\clearpage
%
%
%
%
%
\begin{deluxetable}{ccccccc}
\tablewidth{0pt}
\tablecaption{
Proper Motions, Initial Positions, and Parallaxes of the PZ~Cas H$_2$O 
Maser Spots 
\label{tbl:05}
}
\tablehead{
\colhead{Spot ID           } & 
\colhead{$\Delta\alpha_{A}$} & 
\colhead{$\Delta\delta_{A}$} &
\colhead{\it V$_{\mathrm{LSR}}$} &
\colhead{$\pi$             } &
\colhead{$\mu_{\alpha}
          \cos{\delta}$    } &
\colhead{$\mu_{\delta}$    } \\
\colhead{(Group ID)   } &
\colhead{(mas)             } &
\colhead{(mas)             } &
\colhead{(km~s$^{-1}$)    } &
\colhead{(mas)             } &
\colhead{(mas~yr$^{-1}$)} &
\colhead{(mas~yr$^{-1}$)}
}
\startdata
$1({\it a})$ & $-4.24 \pm 0.07$ & $  52.24 \pm 0.03$ & $-40.9$ & $0.324 \pm 0.042$ & $ -3.79 \pm 0.06$ & $ -1.66 \pm 0.03$ \\
$2({\it a})$ & $-4.20 \pm 0.07$ & $  52.40 \pm 0.04$ & $-41.2$ & $0.379 \pm 0.044$ & $ -3.70 \pm 0.06$ & $ -1.46 \pm 0.03$ \\
$3({\it a})$ & $-4.07 \pm 0.07$ & $  52.70 \pm 0.04$ & $-41.4$ & $0.340 \pm 0.044$ & $ -3.72 \pm 0.06$ & $ -1.30 \pm 0.03$ \\
$4({\it a})$ & $-2.77 \pm 0.07$ & $  56.32 \pm  0.04$ & $-41.4$ & $0.380 \pm 0.039$ & $ -3.67 \pm 0.06$ & $ -1.34 \pm  0.03$ \\
$5({\it a})$ & $-2.83 \pm 0.07$ & $  56.29 \pm  0.03$ & $-41.6$ & $0.357 \pm 0.039$ & $ -3.72 \pm 0.06$ & $ -1.33 \pm  0.03$ \\
$6({\it a})$ & $-2.82 \pm 0.07$ & $  56.28 \pm  0.03$ & $-41.8$ & $0.363 \pm 0.038$ & $ -3.73 \pm 0.06$ & $ -1.30 \pm  0.03$ \\
$7({\it a})$ & $-2.83 \pm 0.07$ & $  56.32 \pm  0.03$ & $-42.0$ & $0.357 \pm 0.038$ & $ -3.76 \pm 0.06$ & $ -1.30 \pm  0.03$ \\
$8({\it a})$ & $-2.90 \pm 0.07$ & $  56.32 \pm  0.03$ & $-42.2$ & $0.355 \pm 0.039$ & $ -3.74 \pm 0.06$ & $ -1.25 \pm  0.03$ \\
$9({\it a})$ & $-2.62 \pm 0.07$ & $  56.62 \pm  0.03$ & $-42.4$ & $0.302 \pm 0.039$ & $ -3.78 \pm 0.06$ & $ -1.28 \pm  0.03$ \\
$10({\it a})$ & $-2.49 \pm 0.07$ & $  56.70 \pm  0.04$ & $-42.6$ & $0.355 \pm 0.046$ & $ -3.95 \pm 0.06$ & $ -1.53 \pm  0.03$ \\
$11({\it a})$ & $-2.39 \pm 0.08$ & $  58.82 \pm  0.05$ & $-42.2$ & $0.436 \pm 0.045$ & $ -3.74 \pm 0.06$ & $ -1.81 \pm  0.03$ \\
$12({\it a})$ & $-2.27 \pm 0.07$ & $  58.65 \pm  0.04$ & $-42.4$ & $0.349 \pm 0.043$ & $ -3.69 \pm 0.06$ & $ -1.82 \pm  0.03$ \\
$13({\it a})$ & $-3.14 \pm 0.07$ & $  61.47 \pm  0.04$ & $-43.1$ & $0.357 \pm 0.051$ & $ -3.81 \pm 0.08$ & $ -1.62 \pm  0.03$ \\
$14({\it b})$ & $63.82 \pm 0.07$ & $  -45.57 \pm  0.03$ & $-40.9$ & $0.367 \pm 0.044$ & $ -2.28 \pm 0.07$ & $ -2.86 \pm  0.03$ \\
$15({\it b})$ & $63.75 \pm 0.07$ & $  -45.49 \pm  0.03$ & $-41.2$ & $0.349 \pm 0.041$ & $ -2.30 \pm 0.07$ & $ -2.85 \pm  0.03$ \\
$16({\it b})$ & $63.70 \pm 0.07$ & $  -45.46 \pm  0.04$ & $-41.4$ & $0.335 \pm 0.044$ & $ -2.32 \pm 0.07$ & $ -2.84 \pm  0.03$ 
\enddata
\end{deluxetable}
\clearpage
%
%
%
\begin{deluxetable}{ccccccc}
\tablewidth{0pt}
\tablecaption{
  Proper Motions and Initial Positions of the Maser Spots  
  Detected in At Least Two Epochs\tablenotemark{a}
  \label{tbl:06}
}
\tablehead{
\colhead{Spot ID           } & 
\colhead{$\Delta\alpha_{A}$} & 
\colhead{$\Delta\delta_{A}$} &
\colhead{V$_{\mathrm{LSR}}$} &
\colhead{$\mu_{\alpha}
          \cos{\delta}$    } &
\colhead{$\mu_{\delta}$    } \\
\colhead{(Group ID)   } &
\colhead{(mas)             } &
\colhead{(mas)             } &
\colhead{(km~s$^{-1}$)    } &
\colhead{(mas~yr$^{-1}$)} &
\colhead{(mas~yr$^{-1}$)}
}
\startdata
$17({\it a})$ & $-2.99 \pm  0.09$ & $  49.94 \pm  0.06$ & $-40.5$ & $ -3.72 \pm  0.24$ & $ -1.63 \pm  0.11$ \\
$18({\it a})$ & $-2.99 \pm  0.09$ & $  49.85 \pm  0.06$ & $-40.7$ & $ -3.60 \pm  0.20$ & $ -1.57 \pm  0.09$ \\
$19({\it a})$ & $-3.02 \pm  0.08$ & $  49.79 \pm  0.08$ & $-40.9$ & $ -3.62 \pm  0.20$ & $ -1.59 \pm  0.10$ \\
$20({\it a})$ & $-3.05 \pm  0.08$ & $  49.83 \pm  0.09$ & $-41.2$ & $ -3.67 \pm  0.20$ & $ -1.64 \pm  0.10$ \\
$21({\it a})$ & $-3.13 \pm  0.09$ & $  49.99 \pm  0.15$ & $-41.4$ & $ -3.78 \pm  0.27$ & $ -0.73 \pm  0.11$ \\
$22({\it a})$ & $-4.36 \pm  0.08$ & $  52.05 \pm  0.05$ & $-40.5$ & $ -3.87 \pm  0.12$ & $ -1.45 \pm  0.06$ \\
$23({\it a})$ & $-4.28 \pm  0.07$ & $  52.09 \pm  0.04$ & $-40.7$ & $ -3.80 \pm  0.07$ & $ -1.72 \pm  0.03$ \\
$24({\it a})$ & $-3.78 \pm  0.08$ & $  52.46 \pm  0.07$ & $-41.6$ & $ -3.81 \pm  0.09$ & $ -1.29 \pm  0.03$ \\
$25({\it a})$ & $-2.66 \pm  0.09$ & $  56.15 \pm  0.06$ & $-41.2$ & $ -3.60 \pm 0.08$ & $ -1.48 \pm  0.04$ \\
$26({\it a})$ & $-1.42 \pm  0.07$ & $  57.46 \pm  0.04$ & $-42.4$ & $ -3.59 \pm 0.42$ & $ -1.79 \pm  0.28$ \\
$27({\it a})$ & $-2.25 \pm  0.07$ & $  56.82 \pm  0.03$ & $-42.2$ & $ -4.13 \pm 0.19$ & $ -2.04 \pm  0.09$ \\
$28({\it a})$ & $-2.19 \pm  0.08$ & $  58.48 \pm  0.05$ & $-42.6$ & $ -3.70 \pm 0.07$ & $ -1.82 \pm  0.03$ \\
$29({\it a})$ & $-3.02 \pm  0.10$ & $  61.42 \pm  0.07$ & $-42.6$ & $ -3.67 \pm 0.13$ & $ -1.70 \pm  0.04$ \\
$30({\it a})$ & $-3.04 \pm  0.09$ & $  61.32 \pm  0.04$ & $-42.8$ & $ -3.84 \pm 0.09$ & $ -1.59 \pm  0.03$ \\
$31({\it a})$ & $-3.08 \pm  0.08$ & $  61.35 \pm  0.04$ & $-43.3$ & $ -3.80 \pm 0.08$ & $ -1.68 \pm  0.03$ \\
$32({\it b})$ & $63.78 \pm  0.09$ & $  -45.68 \pm  0.06$ & $-40.5$ & $ -1.71 \pm 0.17$ & $ -3.08 \pm  0.06$ \\
$33({\it b})$ & $63.92 \pm  0.07$ & $  -45.71 \pm  0.04$ & $-40.7$ & $ -2.26 \pm 0.07$ & $ -2.70 \pm  0.03$ \\
$34({\it b})$ & $63.74 \pm  0.08$ & $  -45.52 \pm  0.05$ & $-41.6$ & $ -2.34 \pm 0.10$ & $ -2.85 \pm  0.04$ \\
$35({\it b})$ & $60.50 \pm  0.10$ & $  -44.46 \pm  0.08$ & $-41.2$ & $ -1.53 \pm 0.70$ & $ -3.30 \pm  0.04$ \\
$36({\it b})$ & $62.97 \pm  0.08$ & $  -43.86 \pm  0.05$ & $-41.4$ & $ -2.70 \pm 0.08$ & $ -2.31 \pm  0.04$ \\
$37({\it b})$ & $63.00 \pm  0.08$ & $  -43.74 \pm  0.06$ & $-41.6$ & $ -2.72 \pm 0.10$ & $ -2.54 \pm  0.04$ \\
$38({\it c})$ & $-31.47 \pm 0.10$ & $ -66.14 \pm  0.17$ & $-42.0$ & $ -4.07 \pm 0.17$ & $ -2.40 \pm 0.05$ \\
$39({\it c})$ & $-31.48 \pm 0.09$ & $ -66.14 \pm  0.10$ & $-42.2$ & $ -4.03 \pm 0.15$ & $ -2.44 \pm 0.05$ \\
$40({\it c})$ & $-33.02 \pm 0.10$ & $ -67.47 \pm  0.07$ & $-42.4$ & $ -4.00 \pm 0.30$ & $ -2.87 \pm 0.06$ \\
$41({\it d})$ & $ 0.03 \pm  0.08$ & $  77.85 \pm  0.06$ & $-48.3$ & $ -1.95 \pm 1.44$ & $ -0.35 \pm  0.17$ \\
$42({\it d})$ & $-0.06 \pm  0.08$ & $  78.28 \pm  0.06$ & $-48.5$ & $ -3.42 \pm 1.37$ & $  0.71 \pm  0.17$ \\
$43({\it g})$ & $-3.97 \pm  0.08$ & $  1.15 \pm  0.05$ & $-41.2$ & $ -3.77 \pm 1.04$ & $ -1.95 \pm  0.11$ \\
$44({\it g})$ & $-3.98 \pm  0.08$ & $  1.06 \pm  0.05$ & $-41.4$ & $ -3.87 \pm 0.27$ & $ -1.97 \pm  0.05$ \\
$45({\it g})$ & $-3.89 \pm  0.09$ & $  1.12 \pm  0.06$ & $-41.6$ & $ -4.00 \pm 0.15$ & $ -1.98 \pm  0.04$ \\
$46({\it g})$ & $-6.51 \pm  0.09$ & $  0.67 \pm  0.07$ & $-41.8$ & $ -2.38 \pm 1.61$ & $ -4.09 \pm  0.05$ \\
\enddata
  \tablenotetext{a}{ 
  Maser spots listed in 
  Table~\ref{tbl:05}
  are excluded. }
\end{deluxetable}
\clearpage
%
%
%
\begin{table}
  \begin{center}
  \caption{Astrometry Analysis Results of PZ~Cas }
  \label{tbl:07}
  \footnotesize
  \begin{tabular}{lcc}
\tableline\tableline
     Annual Parallax   
   & \multicolumn{2}{c}{
           $\pi^{*} =0.356 \pm$ 0.026 mas (2.81$^{+0.22}_{-0.19}$ kpc) } \\
\tableline           
    Position (J2000)\tablenotemark{a}
  & Right Ascension 
  & Declination \\
    (2006 Apr 20)
  & $23^{\mathrm{h}} 44^{\mathrm{m}} 03^{\mathrm{s}}.2816$~$\pm$~0$^s$.0004
  & $+61^{\circ} 47' 22''.187$~$\pm$~0$''$.003 \\
\tableline
    Stellar proper motion 
  & $\mu^{*}_{\alpha}\cos{\delta}=   -3.7 \pm 0.2$~mas~yr$^{-1}$
  & $\mu^{*}_{\delta}=     -2.0 \pm 0.3$~mas~yr$^{-1}$ \\
\tableline\tableline
  \end{tabular}
\tablenotetext{a}{
        The position errors of J2339+6010 are not 
        taken into consideration.  
        This position is located at the origin of the PZ~Cas image 
        of Figure~\ref{fig:05}.}
\end{center}
\end{table}
\clearpage
%
%
%
%
\begin{table}
  \begin{center}
  \caption{Proper Motions of the Member Stars of Cas~OB5 }
  \label{tbl:08}
  \begin{tabular}{lcccccl}
\tableline\tableline
    Object
  & $l$\tablenotemark{a}
  & $b$\tablenotemark{a}
  & $\mu_{\alpha}\cos{\delta}$\tablenotemark{a}
  & $\mu_{\delta}$\tablenotemark{a} 
  & $V$\tablenotemark{b} 
  & Spectl Type\tablenotemark{c} \\
  & (deg) & (deg) & (mas~yr$^{-1}$) & (mas~yr$^{-1}$) & (km~s$^{-1}$) & \\
  \tableline
    HD~223385
  & $115.71$
  & $0.22$
  & $-2.80$
  & $-1.45$
  & $-55.7$\tablenotemark{d}
  & A3~I~A+ \\
    HD~223767
  & $116.02$
  & $-0.20$
  & $-3.51$
  & $-0.92$
  & $-38$\tablenotemark{e}
  & A4~I~AB \\
    HD~223960
  & $115.97$
  & $-1.24$
  & $-2.57$
  & $-2.67$
  & $-48.1$\tablenotemark{f}
  & A0~I~A+ \\
    BD~+61~2550
  & $116.08$
  & $0.04$
  & $-3.22$
  & $-1.25$
  & $-19$\tablenotemark{e}
  & O9.5~II \\
    HD~224424
  & $116.21$
  & $-2.45$
  & $-0.24$
  & $-3.31$
  & $-71$\tablenotemark{f}
  & B1~I~AB \\
    BD~+62~2313
  & $116.31$
  & $1.20$
  & $-2.67$
  & $-2.64$
  & $-53$\tablenotemark{e} 
  & B3~I~B \\
    HD~224055
  & $116.29$
  & $-0.30$
  & $-3.22$
  & $-1.99$
  & $-42.3$\tablenotemark{f} 
  & B1~I~A\\
    LSI~+61~106
  & $116.47$
  & $-0.77$
  & $-3.07$
  & $-1.71$
  & $$
  & B1~V \\
    HD~225146
  & $117.23$
  & $-1.24$
  & $-1.27$
  & $-2.43$
  & $-29.0$\tablenotemark{f} 
  & O9.7~I~B \\
    HD~225094
  & $117.63$
  & $1.26$
  & $-3.27$
  & $-1.83$
  & $-43$\tablenotemark{f} 
  & B3~I~A \\
    HD~108
  & $117.93$
  & $1.25$
  & $-5.39$
  & $-0.93$
  & $-62.8$\tablenotemark{g} 
  & O5F \\
   $\rho$ Cas 
  & $115.30$
  & $-4.52$
  & $-4.54$
  & $-3.45$
  & $-43.1$\tablenotemark{f} 
  & F8~I~ap \\
   KN Cas 
  & $118.15$
  & $0.19$
  & $-2.50$
  & $2.02$
  & $-87.3$\tablenotemark{h} 
  & M1~I~bep+B \\
    TZ~Cas
  & $115.90$
  & $-1.07$
  & $-3.05$
  & $-3.56$
  & $-54.28$\tablenotemark{i} 
  & M2~I~ab \\
    PZ~Cas
  & $115.06$
  & $-0.05$
  & $-3.52$
  & $-3.55$
  & $-45.68$\tablenotemark{i} 
  & M4~I~a \\
\tableline\tableline
  \end{tabular}
   \tablenotetext{a}{
  \cite{Leeuwen2007}. }
  \tablenotetext{b}{
  Radial velocity (heliocentric coordinate)}
   \tablenotetext{c}{
  \cite{Garmany1992}. }
  \tablenotetext{d}{ 
  \cite{Valdes2004}.} 
  \tablenotetext{e}{ 
  \cite{Fehrenbach1996}.}
   \tablenotetext{f}{
   \cite{Wilson1953}.}
   \tablenotetext{g}{
   \cite{Evans1967}.}
   \tablenotetext{h}{
   \cite{Humphreys1970}.}
   \tablenotetext{i}{
   \cite{Famaey2005}.}
\end{center}
\end{table}
\clearpage
%
%
\begin{table}
  \begin{center}
  \caption{Proper Motions of the Member Stars of Cas~OB4 and Cas~OB7 }
  \label{tbl:09}
  \begin{tabular}{lcccccl}
\tableline\tableline
    Object
  & $l$\tablenotemark{a}
  & $b$\tablenotemark{a}
  & $\mu_{\alpha}\cos{\delta}$\tablenotemark{a}
  & $\mu_{\delta}$\tablenotemark{a} 
  & $V$\tablenotemark{b} 
  & Spectl Type\tablenotemark{c} \\
  & (deg) & (deg) & (mas~yr$^{-1}$) & (mas~yr$^{-1}$) & (km~s$^{-1}$) & \\
  \tableline
    Cas~OB4
  & $$
  & $$
  & $$
  & $$
  & $$
  & \\
  \tableline
    HD~1544
  & $119.27$
  & $-0.58$
  & $-1.38$
  & $-1.11$
  & $-52.0$\tablenotemark{d}
  & B0.5~III \\
    HD~2451
  & $120.32$
  & $-0.25$
  & $-1.22$
  & $-2.11$
  & $-37$\tablenotemark{d}
  & B0.5~IV \\
    HILTNER~8
  & $117.69$
  & $-4.25$
  & $0.25$
  & $-2.49$
  & $-23.5$\tablenotemark{e}
  & O9~IV \\
    DL~Cas
  & $120.27$
  & $-2.55$
  & $-0.75$
  & $-1.38$
  & $-36.80$\tablenotemark{f}
  & F7.5~I~B \\
    HD~236433
  & $120.28$
  & $-2.61$
  & $-2.02$
  & $-2.12$
  & $-11.2$\tablenotemark{g}
  & F5~I~B \\
  \tableline
   Cas~OB7
  & $$
  & $$
  & $$
  & $$
  & $$
  & \\
  \tableline
    HD~3940
  & $122.00$
  & $1.44$
  & $-2.61$
  & $-0.69$
  & $-48.0$\tablenotemark{h} 
  & A1~I~A \\
    HD~4694
  & $122.77$
  & $1.77$
  & $-0.90$
  & $0.04$
  & $-49.0$\tablenotemark{i} 
  & B3~I~A \\
    HD~4841
  & $122.93$
  & $0.91$
  & $-1.37$
  & $-0.93$
  & $-26$\tablenotemark{e} 
  & B5~I~A \\
    HD~5551
  & $123.71$
  & $0.85$
  & $-2.05$
  & $-1.53$
  & $-51.0$\tablenotemark{d} 
  & B1.5~I~A \\
    HD~4842
  & $122.93$
  & $0.05$
  & $-28.98$
  & $-2.81$
  & $-92$\tablenotemark{e} 
  & M \\
    HD~6474
  & $124.65$
  & $0.95$
  & $-1.17$
  & $-1.00$
  & $-48.3$\tablenotemark{j} 
  & G4~I~A \\
\tableline\tableline
  \end{tabular}
    \tablenotetext{a}{
  \cite{Leeuwen2007}. }
   \tablenotetext{b}{
   Radial velocity (heliocentric coordinate)}
   \tablenotetext{c}{
  \cite{Garmany1992}. }
   \tablenotetext{d}{
   \cite{Evans1967}.}  
   \tablenotetext{e}{
   \cite{Wilson1953}.}
  \tablenotetext{f}{ 
  \cite{Mermilliod2008}.} 
  \tablenotetext{g}{ 
  \cite{Kraft1958}.}
   \tablenotetext{h}{
   \cite{Munch1957}.}  
    \tablenotetext{i}{
  \cite{Abt1972}. }
   \tablenotetext{j}{
   \cite{Humphreys1970}.}
\end{center}
\end{table}
\clearpage
%
%



\clearpage




\end{document}